\documentclass[twocolumn]{emulateapj}
\usepackage{amstext}
\usepackage{amsmath}
\usepackage{amssymb}
\usepackage{xcolor,xspace}
\usepackage{multirow,threeparttable,tabularx}
\usepackage{aas_macros}
\usepackage[colorlinks=true,linkcolor=blue,citecolor=blue]{hyperref}

\newcommand{\sa}{Sgr~A*\xspace}
\newcommand{\se}{Sgr~A~East\xspace}

\begin{document}

\shorttitle{Debris Streams and Remnants Resulting from Tidal Disruptions}

\shortauthors{Guillochon et al.}

\title{Unbound Debris Streams and Remnants Resulting From the Tidal Disruptions of Stars by Supermassive Black Holes}

\author{James Guillochon\altaffilmark{1,2}, Michael McCourt\altaffilmark{1,3}, Xian Chen\altaffilmark{4}, Michael D. Johnson\altaffilmark{1}, Edo Berger\altaffilmark{1}}
\altaffiltext{1}{Harvard-Smithsonian Center for Astrophysics, The Institute for Theory and
Computation, 60 Garden Street, Cambridge, MA 02138, USA}
\altaffiltext{2}{Einstein Fellow}
\altaffiltext{3}{ITC Fellow}
\altaffiltext{4}{Instituto de Astrof\'{i}sica, Facultad de F\'{i}sica, Pontificia Universidad Cat\'{o}lica de Chile, 782-0436 Santiago, Chile}

\email{jguillochon@cfa.harvard.edu}

\begin{abstract} 
The kinetic energy of a star in orbit about a supermassive black hole is a significant fraction of its rest mass energy when its periapse is comparable to its tidal radius. Upon its destruction, a fraction of this energy is extracted and injected into the stellar debris, half of which becomes unbound from the black hole, with the fastest material moving at $\sim 0.03 c$. In this paper, we present a formalism for determining the fate of these unbound debris streams (UDSs) as they depart from the black hole and interact with the surrounding gas. As the density and velocity varies along the length of a UDS, we find that hydrodynamical drag quickly shapes UDSs into loop-like structures, with the densest portions of the streams leading portions of lower density. As UDSs travel outwards, their drag against the ISM increases quadratically with distance, which causes UDSs to deposit their momentum and energy into the ambient medium before the surrounding shocked ISM has a chance to cool. This sudden injection of $\sim 10^{50}$ erg into the ambient medium generates a Sedov-like unbound debris remnant (UDR) that mimics supernova remnants (SNRs) in energetics and appearance, accelerates particles which will produce cosmic rays and synchrotron emission, and provides momentum feedback into the molecular clouds surrounding a black hole. We estimate that a few of these UDRs might be present within a couple degrees of the Galactic Center masquerading as SNRs, and that the UDR scenario is a plausible explanation for \se.
\end{abstract}


\section{Introduction}
The full or partial tidal disruption of a star by a supermassive black hole results in two streams of debris: A bound stream that falls into the black hole and powers a luminous flare, and an unbound stream that is launched from the black hole at a velocity larger than the black hole's escape velocity \citep{Rees:1988a}. Because the accretion of the bound stream produces AGN-like luminosities, it has been the focus of most studies on tidal disruption events (TDEs), with many flares purportedly originating from its accretion\footnote{See the Catalogue of Possible Tidal Disruption Events for an up to date listing, \url{http://astrocrash.net/resources/tde-catalogue}} \citep{Komossa:2004a,Gezari:2012a,Arcavi:2014a}. In this paper we focus upon the unbound debris stream (UDS) and its interaction with the interstellar medium that surrounds the disrupting black hole, along with the observational signatures of this interaction.

Observations suggest that the observed rate of luminous flares associated with the accretion of bound debris \citep[$10^{-5}$~yr$^{-1}$,][]{Cenko:2012b,van-Velzen:2014a,Stone:2014a,Holoien:2015a} is 10\% the theoretical disruption rate of $10^{-4}$~yr$^{-1}$ \citep{Merritt:2010c}. This deficit is possibly a result of inefficient accretion about many black holes when relativistic effects are weak \citep{Guillochon:2015b,Shiokawa:2015a,Bonnerot:2015a,Hayasaki:2015a}. This means that the true rate of tidal disruptions may be close to the theoretically expected value, but with only a small fraction producing rapidly-evolving flares that would be identified as TDEs. However, {\it all} disruptions, whether they produce a luminous flare or not, will eject $\sim$50\% of the star's mass into a UDS. The center of the Milky Way hosts a number of supernova remnants (SNRs), with approximately a half-dozen lying within a few degrees of the central black hole \sa \citep{LaRosa:2000a}. Given the average SNR lifetime of $10^{4}$~yr, this implies a supernova (SNe) rate of a few times $10^{-4}$~yr$^{-1}$; the similarity between this rate and the TDE rate suggests that SNe and TDEs may both be relevant for shaping and heating the gas in the centers of galaxies.

In this paper, we show that each UDS deposits an energy and momentum that are both on average an order of magnitude smaller than that injected by a single supernova, with $\sim 10^{50}$~erg and $\sim 10^{41}$~g~cm~s$^{-1}$ being typical. Because the disruption rate is comparable to the supernova rate within the same volume, and because remnant lifetimes are only mildly sensitive to the injected energy \citep{Blondin:1998a}, this implies that a few unbound debris remnants (UDRs) are potentially visible in the Galactic Center (GC) region. \se is a remnant located several pc distant from \sa which had previously been suggested as a UDR \citep{Khokhlov:1996a}, however, observational evidence of the total energy content of \se \citep{Park:2005a} argued that a SNR was more likely. We argue that this disagreement was mostly due to a theoretical overestimate in the amount of energy deposited by UDSs, and we show that UDRs in fact have excellent quantitative agreement with many of \se's observed properties.

In Section \ref{sec:dynamics} we describe the basics of UDS evolution and calculate its energetics. In Section \ref{sec:method} we present a method for calculating the temporal evolution of UDSs after they are produced by a stellar disruption, and describe how we select our input parameters. Section \ref{sec:uds} presents the results of our numerical approach, shows how the UDSs evolve, and how they deposit energy and momentum into the environment. The UDRs that we propose are produced when UDSs decelerate are characterized in Section \ref{sec:udr}. The impact of UDSs and UDRs are considered in Section \ref{sec:impact}, where we explore the consequences of the energy and momentum injected by them, including their contribution to cosmic rays, radio emission, and feedback into surrounding molecular clouds. We summarize our results in Section \ref{sec:discussion}.

\section{Dynamics of the Unbound Debris}\label{sec:dynamics}
All parts of the star travel away from the black hole post-disruption, but for the bound stream, each parcel of matter turns around at an apoapse defined by its binding energy to the black hole, which ranges from zero to $-\Delta \epsilon \sim -q^{1/3} v_{\ast}^{2}$, where $q \equiv M_{\rm h}/M_{\ast}$, $M_{\rm h}$ is the black hole mass, $v_{\ast} \equiv \sqrt{2 G M_{\ast} / R_{\ast}}$ is the escape velocity from the surface of the star, and $M_{\ast}$ and $R_{\ast}$ are the star's mass and radius \citep{Rees:1988a}. Because of the large mass ratio associated with tidal disruptions by supermassive black holes, the tidal force differs little between the near- and far-sides of the star \citep{Guillochon:2011a}. As a result, the unbound stream has a nearly identical spread in energy to the bound stream, except that each parcel of gas has positive kinetic energy ranging from $0$ to $+\Delta \epsilon$. This corresponds to a maximum velocity $v_{\max} \sim q^{1/6} v_{\ast}$, which when canonicalized to a solar-type star about our own galaxy's supermassive black hole is
\begin{align}
v_{\max} &= {\rm 8,\!000} \left(\frac{M_{\ast}}{M_{\odot}}\right)^{1/3} \left(\frac{R_{\ast}}{R_{\odot}}\right)^{-1/2}\nonumber\\
&\times \left(\frac{M_{\rm h}}{4 \times 10^{6} M_{\odot}}\right)^{1/6} {\rm km~s}^{-1},\label{eq:vmax}
\end{align}
which, if not slowed by drag, could easily escape the galaxy \citep{Kenyon:2008a}. If all parts of a UDS possessed this velocity, a solar mass disruption would yield $10^{51}$~erg of kinetic energy \citep{Cheng:2011a}, but because the unbound debris has a range of velocities, with a large fraction of the mass moving significantly slower, the typical kinetic energies end up being closer to $10^{50}$~erg per event.

As the stream expands, it remains self-gravitating for a period of time, restricting the growth of its width significantly \citep{Kochanek:1994a,Guillochon:2014a,Coughlin:2015a}, with the cylindrical radius of the stream $s \propto r^{1/4}$, where $r$ is the distance from the black hole. Once the stream temperature drops below several $10^{3}$~K, it begins to recombine \citep{Roos:1992a,Kochanek:1994a,Kasen:2010a}; this recombination injects heat into the gas which causes it to expand. Once the density of the stream drops by a factor of a few, the stream no longer satisfies the Jeans condition and ceases to be self-gravitating, expanding self-similarly with $s \propto r$.

Except for debris that is marginally unbound to the hole (a small fraction of the total mass), most of the unbound debris is on hyperbolic trajectories and travels at a velocity directly proportional to its distance to the black hole. Because the initial velocity increases monotonically along the stream's length, the stream continually stretches in the radial direction. At the same time, the stream expands laterally at its internal sound speed $c_{\rm s}$ after it has broken free from its own gravity. The resulting expansion is thus homologous with the stream's density dropping as $t^{-3}$, and thus we would expect the stream's density to eventually reach densities comparable to the background ISM density. For a MS star, the initial average density is $\sim 10^{24}$ times larger than the ISM density (1~g~cm$^{-3}$ vs. 1 particle per cm$^{-3}$), and so even ignoring self-gravity and assuming homologous expansion from the tidal radius $r_{\rm t} \equiv (M_{\rm h} / M_{\ast})^{1/3} R_{\ast}$ yields equal stream and background densities at a distance of $r_{\rm t} (10^{24})^{1/3} \simeq 1$~kpc, with a travel time of $\sim 10^{6}$~yr. Giant stars, which can be $\sim 10^{-3}$ times less dense than MS stars, would reach equal densities at distances closer to 100~pc, with a $10^{5}$~yr travel time.

The stream's terminal distance is determined by when the amount of mass it sweeps up is comparable to its own mass. If the constituent parts of the stream traveled radially outward through a near-constant density ISM, this stalling distance $r_{\rm stall}$ would be similar to the distance at which the stream density becomes comparable to the background. In fact, the UDSs stop much faster than this simple estimate suggests. The stream is ejected with a range of velocities, and each piece of the stream is on a slightly different hyperbolic orbit. As the unbound stream must connect to the bound stream at $\epsilon = 0$, the stream forms an arc which spans an angle $\theta_{\rm arc}$, where $\theta_{\rm arc}$ is measured from the argument of apoapse of the $\epsilon \sim 0$ debris. The angle $\theta_{\rm arc}$ is defined by the hyperbolic angle of the least bound material,
\begin{equation}
\theta_{\rm arc} = \arccos \left[\left(q^{-1/3}+1\right)^{-1}\right] \simeq \sqrt{2} q^{-1/6}.\label{eq:thetaarc}
\end{equation}
This results in an increase of the effective cross section of the stream over the straight-path case of $\theta_{\rm arc} / \theta_{\ast} = \sqrt{2} q^{1/6}$, where $\theta_{\ast} = R_{\ast} / (\pi r_{\rm t})$ is the angular size of the star at the tidal radius. In our own GC, the disruption of a solar-mass star ($q = 4 \times 10^{6}$) would yield $\theta_{\rm arc} / \theta_{\ast} \simeq 20$. This suggests that the true stalling distance $r_{\rm stall} \simeq 50$~pc (5~pc for giants) rather than 1~kpc. Note that even for giant disruptions that this distance is greater than $r_{\rm h}$, and that the debris mass represents only a small fraction of the total mass contained within that region. Thus, it is unlikely that the deposited debris would affect the global flow structure surrounding the black hole, as is suggested by \citet{Gopal-Krishna:2008a}.

This estimate still approximates the dynamics of UDSs in that it presumes that the background density is constant as a function of $r$, and that the stream's mass is evenly spread in energy from $0$ to $\Delta \epsilon$. In reality, the spread of stellar debris as a function of binding energy is non-trivial and depends on the star's structure \citep{Lodato:2009a} and the proximity of the star to the black hole at periapse \citep{Guillochon:2013a}. This motivates the quantitative work we describe in the next section.
\bigskip\bigskip

\section{Method}\label{sec:method}
To track the evolution of UDSs after being produced shortly after periapse of the progenitor star, we discretize the star into $N$ segments of equal mass $\delta m_{i} = M_{\ast}/N$ ($i \in \left\{1, 2, \ldots, N\right\}$) and consider the forces on each mass
element. For each mass element $\delta m_{i}$, we determine its initial binding energy $\epsilon_{i}$ using the distributions of mass as a function of energy $dM/dE$ from \citet{Guillochon:2013a}:
\begin{align}
  \epsilon_{i} = \mathcal{M}^{-1}\left[\left(i-\frac{1}{2}\right) \delta{}m_{i}\right],
\end{align}
where $\mathcal{M}(\epsilon) \equiv \int_0^{\epsilon} (dM/dE)dE$ is
the cumulative distribution of the unbound mass.  Using this binding
energy, we specify initial positions ${\bf x}_{i}$ and velocities
$\dot{\bf x}_{i}$ for each segment $i$ at $t = 0$,
\begin{align}
{\bf x}_{i}(0) &= \left[r_{\rm p} + R_{\ast}\left(\frac{\epsilon_{i}}{\epsilon_{\max}}\right)\right] \hat{x}\\
{\bf \dot{x}}_{i}(0) &= \sqrt{2\left[\frac{G M_{\rm h}}{\left|{\bf x}_{i}(0)\right|}+\epsilon_{i}\right]} \hat{y},
\end{align}
where $r_{\rm p}$ is the star's original periapse distance. Since we have restricted our analysis to the unbound debris with
$\epsilon_{i}>0$, this is an initially hyperbolic trajectory; we choose
our coordinate system such that this trajectory lies in the
$x$--$y$~plane.  Each segment subsequently obeys an equation of
motion,
\begin{equation}
\ddot{\bf x}_{i} = \boldsymbol{f}_{{\rm g},i} + \boldsymbol{f}_{{\rm d},i}
\label{eq:xddot},
\end{equation}
where $\boldsymbol{f}_{{\rm g},i}$ and $\boldsymbol{f}_{{\rm d},i}$ are the gravitational and drag forces per unit mass (acceleration) applied to the segment, respectively.

\begin{figure}
\centering\includegraphics[width=0.95\linewidth,clip=true]{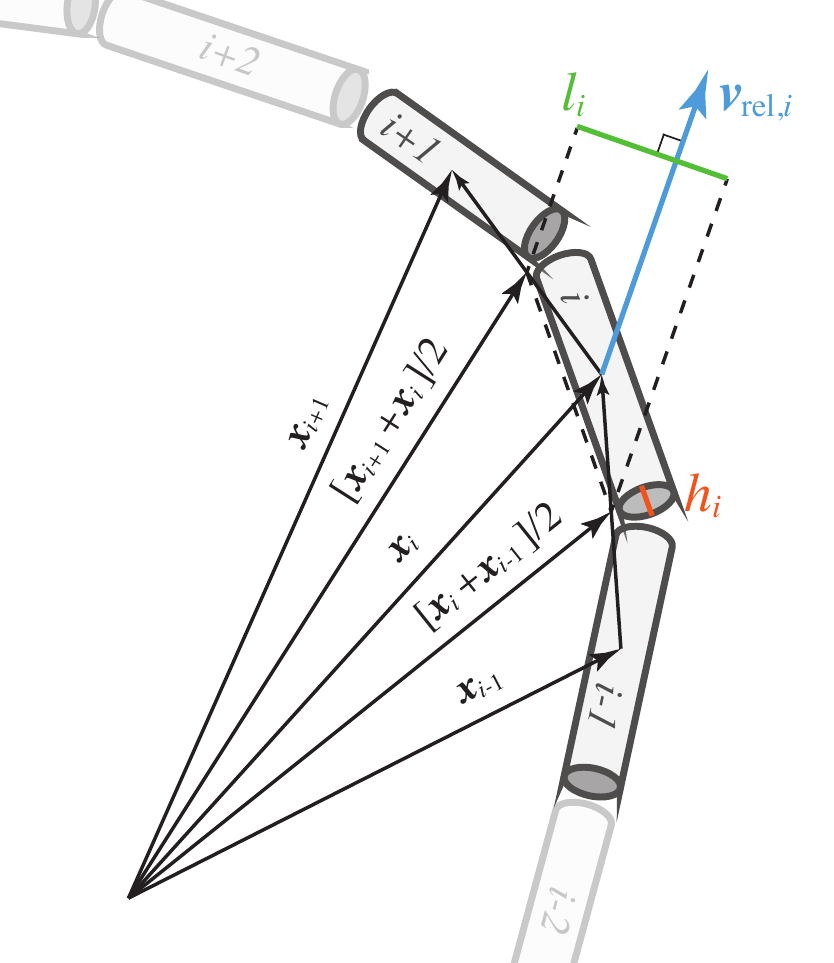}
\caption{Diagram demonstrating how the projected length $l_{i}$ (shown in green) and thickness $h_{i}$ (shown in orange) are determined for the $i$th segment in our formalism given its position ${\bf x}_{i}$, the position of the neighboring segments ${\bf x}_{i-1}$ and ${\bf x}_{i+1}$, and the relative velocity vector $v_{{\rm rel},i}$ (shown in blue). For illustrative purposes, the thickness $h_{i}$ shown is greatly exaggerated relative to the typical $l_{i}$.}
\label{fig:diagram}
\end{figure}

Prior to disruption, the star is approximately spherical in shape. Because the drag forces are entirely negligible shortly after disruption (as the density contrast is $\sim 10^{20}$), each mass element in the star initially follows a free, hyperbolic trajectory; within a few dynamical times, the star therefore expands into a highly elongated, spaghetti-like structure.  The resulting stream is approximately circular in cross-section, with a cylindrical radius $s$ which is much less than the local radius of curvature of the stream ($\sim r$).  The mass elements $\delta{}m_{i}$ can thus be well-approximated as a discrete set of connected cylinders, as depicted in Figure~\ref{fig:diagram}.  We determine the center of mass of each cylinder using Equation~\ref{eq:xddot}, and the orientation using neighboring segments as depicted in the diagram.

The gravitational force per unit mass $\boldsymbol{f}_{{\rm g},i}$ in Equation \ref{eq:xddot} originates from two components,
\begin{align}
\boldsymbol{f}_{{\rm g},i} &= -G \left[\frac{M_{\rm h}}{\left|{\bf x}_{i}\right|^{3}} + \frac{M_{\rm b}}{\left|{\bf x}_{i}\right|\left(\left|{\bf x}_{i}\right| - r_{\rm b}\right)^{2}}\right] {\bf x}_{i}\label{eq:fgrav}
\end{align}
with the first term in the brackets coming from the black hole, and the second coming from the bulge of stars surrounding the black hole, where the mass of the bulge $M_{\rm b} = 3 \times 10^{9} M_{\odot}$ and effective radius of the bulge $r_{\rm b} = 100$~pc are assigned the values suggested in \citet{Kenyon:2008a}.

As each mass element moves through the background gas with density $\rho_{\rm bg}$ and velocity ${\bf v}_{\rm bg}$, it feels a drag force per unit mass
\begin{equation}
\boldsymbol{f}_{{\rm d,i}} = -C \left[\frac{h_{i} l_{i} \rho_{\rm bg}}{\delta m_{i}}\left|{\bf v}_{{\rm rel},i}\right|\right] {\bf v}_{{\rm rel},i}\label{eq:fdrag}
\end{equation}
where the relative velocity ${\bf v}_{{\rm rel},i} = |\dot{\bf x}_{i}| - {\bf v}_{\rm bg}$, the scalars $h_{i}$ and $l_{i}$ represent each segment's vertical and horizontal lengths projected perpendicular to the segment's relative velocity against the background, and $C$ is a drag coefficient which we have set to unity. As depicted in Figure~\ref{fig:diagram}, we approximate the projected length $l_{i}$ as
\begin{equation}
l_{i} = \left|\dfrac{1}{2}\left({\bf x}_{\min\left(i+1,N\right)} - {\bf x}_{\max\left(i-1,1\right)}\right) \times \frac{{\bf v}_{{\rm rel},i}}{\left|{\bf v}_{{\rm rel},i}\right|} \right|.\label{eq:length}
\end{equation}

The thickness of the stream is initially determined by self-gravity, but then is governed by free-expansion,
\begin{equation}
h_{i} = 2 R_{\ast} \left[\frac{\min\left(\left|{\bf x}_{i}\right|, r_{\rm rec}\right)}{r_{\rm p}}\right]^{1/4} + \max\left[c_{\rm s,rec} \left(t - t_{\rm rec}\right), 0\right]\label{eq:thickness},
\end{equation}
where we have made the assumption that the stream recombines and breaks self-gravity at a particular distance $r_{\rm rec}$; the time $t_{\rm rec}$ is approximately when this recombination occurs, and $c_{\rm s,rec} = \sqrt{3 \mu k_{\rm b} T_{\rm rec}/m_{\rm p}}$ is the sound speed at the time of recombination, where $T_{\rm rec}$ is the recombination temperature and $\mu$ the mean molecular weight. In reality, each segment recombines at a slightly different distance and time as each segment has a different density; to compare against our single-value assumption we have performed some numerical tests in which we calculate when each segment recombines by determining when its internal temperature $T_{i}$ drops below $T_{\rm rec}$. As the stream is initially very optically thick \citep{Kasen:2010a}, we assume that $T$ evolves adiabatically before recombination. We also assume that the stream has not yet slowed due to drag forces, a safe assumption given that the density contrast evolves little while the stream remains self-gravitating. The time of recombination can then be determined by when the following condition is met,
\begin{equation}
\left(\frac{h_{i}}{2 R_{\ast}}\right)^{2} \frac{l_{i}}{2 R_{\ast} \delta v_{i}} > \left(\frac{T_{i}}{T_{\rm rec}}\right)^{3/2},
\end{equation}
where $\delta v_{i}$ is approximated as $(v_{i+1} - v_{i-1})/2$. We find that recombination occurs at distances of $10^{14}$ -- $10^{16}$~cm for UDS, corresponding to a time of recombination $10^{4}$ -- $10^{6}$~s. As we will show in Section \ref{sec:uds}, we find that the typical stopping distances for UDSs are $\gtrsim 10^{18}$~cm, even for very dense background conditions, with drag only playing a significant role in UDS evolution at these distances. When $r \gg r_{\rm rec}$, the second term in Equation~(\ref{eq:thickness}) dominates, and $c_{\rm s,rec} (t - t_{\rm rec}) \simeq  c_{\rm s,rec} t$, and thus $h_{i} \simeq c_{\rm s,rec} t$. Thus our particular choice of $r_{\rm rec}$ and $t_{\rm rec}$ do not have much quantitative effect on our results, and for simplicity we set these to constants, $r_{\rm rec} = 10^{15}$~cm and $t_{\rm rec} = 10^{5}$~s.

\subsection{Initial Conditions for a Monte Carlo Exploration of Parameter Space}
We generate a Monte Carlo ensemble of UDSs to produce a statistical population with realistic event-specific parameters. In the present work, we restrict our analysis to a single black hole mass of $4 \times 10^{\smash{6}} M_{\odot}$, approximately equal to recent estimates for the mass of \sa \citep{Ghez:2008a,Gillessen:2009b}, although the latest estimates have sided towards slightly larger values \citep{Do:2013b}.

We use two different cases for the background density. In the first (case~A), we use a background density that scales inversely with distance \citep{Yuan:2003a},
\begin{equation}
\rho_{\rm bg,A} = \rho_{\rm bg, 0} \left[\frac{\left|{\bf x}_{i}\right|}{r_{\rm bg, 0}}\right]^{-1} + \rho_{\rm floor}\label{eq:rhobga},
\end{equation}
where we set the constants $\rho_{\rm bg, 0} = 1.3 \times 10^{-21}$~g~cm$^{-3}$ and $r_{\rm bg, 0} = 1.3 \times 10^{16}$~cm, values similar to those determined from X-ray measurements of Bremsstrahlung emission near the Bondi radius \citep{Quataert:2002a}. A floor value of $\rho_{\rm floor} = 1.7~\times~10^{-24}$~g~cm$^{-3}$ is set to the average Milky Way ISM value.

Case~A is appropriate for a UDS that interacts with the low-density ISM only, but the GC region is partially filled with dense molecular clouds with number densities in excess of $10^{4}$~cm$^{-3}$ \citep{Genzel:2010b}. Radio observations of the GC revealed tens of molecular clouds at a distance of $0.5-2$ pc from the SMBH, together with a large amount of diffusive molecular gas between these clouds \citep{Mezger:1996a}. Most of them reside in a torus-like structure, known as the circum-nuclear disk (CND), which has a half-opening angle of about $27^\circ$ and covers $\sim40\%$ of the sky when viewed from \sa \citep{Christopher:2005a}. The typical density of a MC in the Galactic plane is $10^4$~cm$^{-3}$; however, it is possible that the MCs in the Galactic Center are much denser because of the existence of strong external pressure \citep{Chen:2015a}. To treat the UDS evolution in such regions we define a second density profile (case~B) in which an additional ad-hoc wall of dense material is placed at the black hole's sphere of influence,
\begin{align}
\rho_{\rm bg,B} &= \rho_{\rm bg,A}\nonumber\\
&+\frac{n_{\rm wall}m_{\rm p}}{2} \left(1+\mathop{\rm tanh} \left[10 \left(\frac{\left|{\bf x}_{i}\right| - r_{\rm h}}{r_{\rm h}}\right)\right]\right),\label{eq:rhobgb}
\end{align}
where $n_{\rm wall} = 10^{4}$~cm$^{-3}$ and $r_{\rm h} = 3$~pc. Note that both cases~A and B are different from what is assumed by \citet{Khokhlov:1996a} who set $n = 10^{4}$~cm$^{-3}$ at all radii, whereas observations suggest that such densities are only realized at distances $\lesssim 10^{16}$~cm from \sa or within the molecular clouds in the CND.

\begin{figure*}
\centering\includegraphics[width=0.9\linewidth,clip=true]{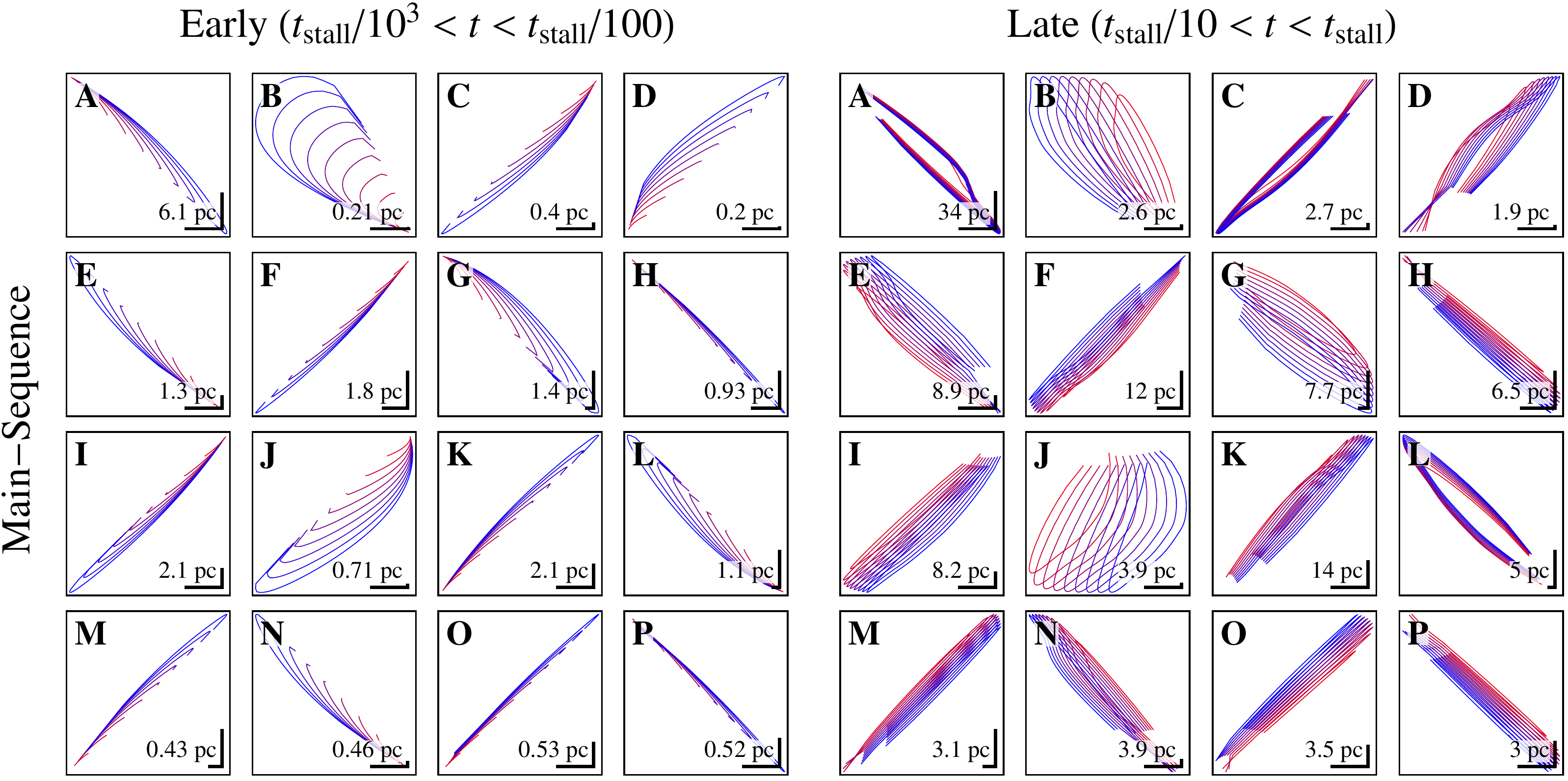}
\caption{Two dimensional projections of UDS evolution originating from the disruptions of MS stars under the case~A assumption, where the image plane is perpendicular to the axis of rotation of the background flow. The figure shows two separate sets of columns that show the early- and late-time evolution of the UDS, where the letter labels uniquely identify each run in both sets (e.g. the two ``F'' panels correspond to the same simulation at early and late times). The left columns show the UDS at early times, with the ten curves in each panel showing the UDS state at time $t$ where $t$ is drawn from evenly-spaced intervals from $t_{\rm stall}/10^{3}$ (red) to $t_{\rm stall}/100$ (blue). The right columns show the UDS at late times, $t$ in these columns ranges from $t_{\rm stall}/10$ (red) to $t_{\rm stall}$ (blue). For visibility, the true stream widths are not shown in this figure. The aspect ratio of each panel is set to unity to make the structures more apparent, but in reality the streams are highly elongated and have large aspect ratios. The black line segments in the lower right corners of each panel show the linear scale of the panel and indicate the true aspect ratio; both segments show the same denoted distance in the $x$ and $y$ directions (e.g. panel ``L'' shows a UDS that is highly elongated in $y$, whereas panel ``B'' shows a UDS that is highly elongated in $x$).}
\label{fig:mstails}
\end{figure*}

\begin{figure*}
\centering\includegraphics[width=0.9\linewidth,clip=true]{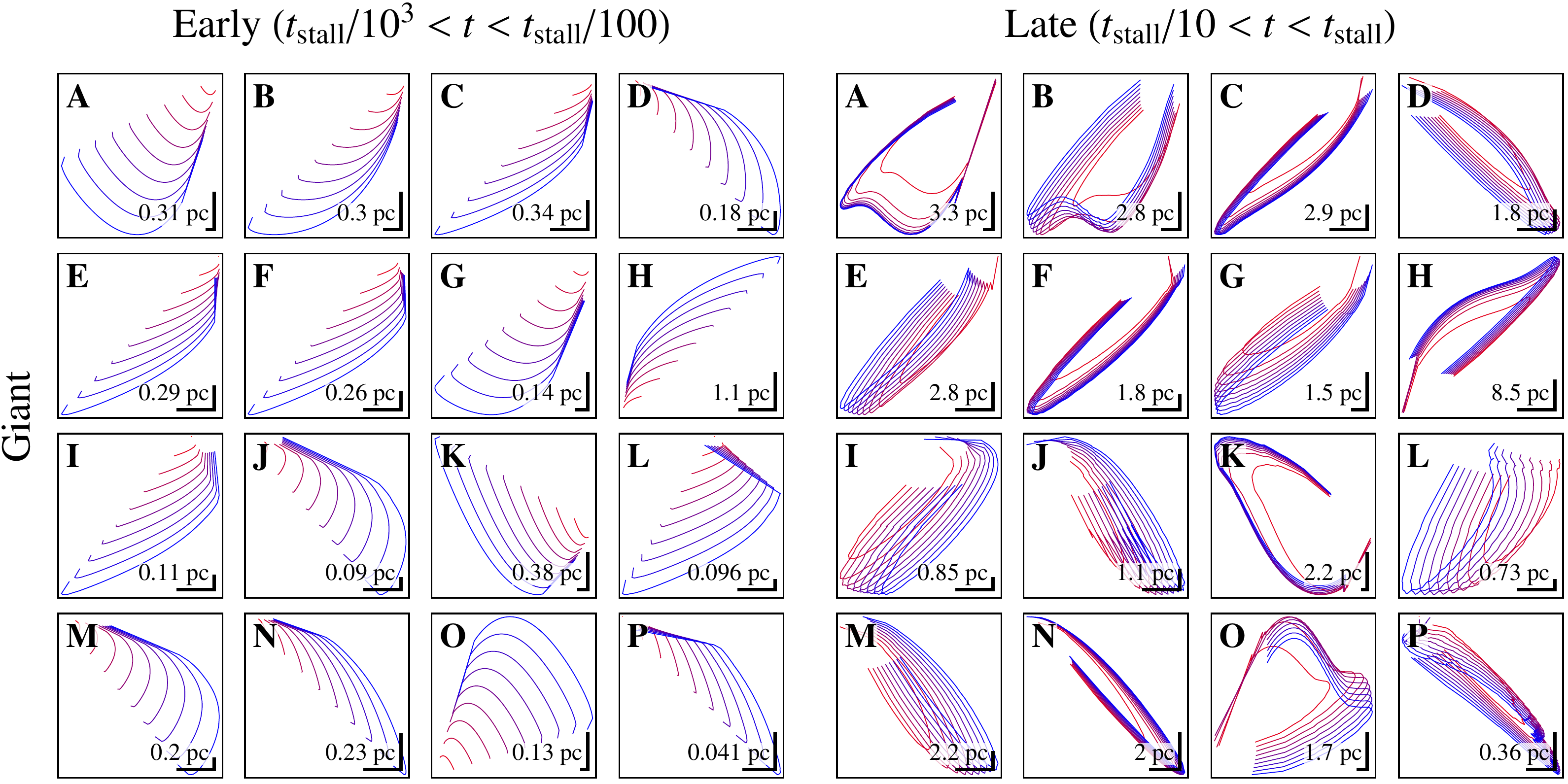}
\caption{Identical to Figure~\ref{fig:mstails}, but for a selection of UDSs generated by the disruption of giant stars.}
\label{fig:gianttails}
\end{figure*}

We assume the ISM is partly rotationally supported as in \citet{McCourt:2015a},
\begin{align}
{\bf v}_{\rm bg} &= f_{\rm kep} \sqrt{\left|{\boldsymbol{f}}_{\rm g}\right|} \frac{{\bf J} \times {\bf x}_{i}}{\left| {\bf J}\right| \left|{\bf x}_{i}\right|}\label{eq:vbg},
\end{align}
where $f_{\rm kep}$ is defined as the background flow velocity's fraction of the local circular velocity in the equatorial plane, $f_{\rm kep} \equiv v_{\rm bg} / v_{\rm circ}$, and for this paper we set $f_{\rm kep} = 0.5$.

We consider two kinds of stars in this study: Main-sequence (MS) stars and giant stars. In combination with the case~A and B background densities defined in Equations (\ref{eq:rhobga}) and (\ref{eq:rhobgb}), the two stellar types give us four combinations of parameters for which we perform independent Monte Carlo calculations: MS (case~A and B), and giants (case~A and B). For each UDS, we vary the impact parameter $\beta$, the mass of the star $M_{\ast}$, and the angular momentum vector describing the background flow {\bf J}, which defines ${\bf v}_{\rm bg}$ (Equation~(\ref{eq:vbg})).

Main-sequence stars are presumed to follow a Kroupa IMF \citep{Kroupa:2001a}, with radii determined by the fitting relations of \citet{Tout:1996a} where we have assumed solar metallicity. Giant stars are drawn from the same Kroupa distribution, approximately the distribution they should follow if star formation is constant and if the giant phase is much shorter than the MS lifetime. Giants are presumed to have a fixed radius of $R_{\ast} = 0.5$~AU.

Because the tidal radii of MS stars are small, they are most likely to be in the ``pinhole'' regime \citep{Lightman:1977a}, resulting in a $\beta$ probability distribution $P(\beta) \propto \beta^{\smash{-2}}$. Giant stars, which are typically much larger than MS stars, are more likely to be in the ``diffusion'' regime, resulting in encounters that typically only graze the tidal radius, with stellar evolution playing an important role \citep{MacLeod:2012a,MacLeod:2013a}. Because multiple encounters are likely for giant stars, we make an ad-hoc choice in the $\beta$ distribution of $P(\beta) \propto \exp (-5 \beta)$ as is appropriate for a diffusion process where the typical change in angular momentum per scattering event is 20\%. As a consequence of this assumption, the vast majority of giant disruption events (98\%) are partial disruptions.

We assume that the orbital planes of disrupted stars are isotropically distributed; since we chose our coordinate system such that the star's orbit always lies in the $x$--$y$ plane, the rotation axis ${\bf J}/|{\bf J}|$ is randomly drawn from the 2-sphere.  Before plotting our results, we rotate the coordinate system such that the background rotation axis is in the $\hat{z}$ direction; when plotting multiple events (e.g. Figure~\ref{fig:streamsgc}) we perform this rotation for each individual event before combining them.

\section{Unbound Debris Streams}\label{sec:uds}

An ensemble of UDSs are shown in Figures \ref{fig:mstails} (for MS stars) and \ref{fig:gianttails} (for giants) using our case~A density profile (Equation~(\ref{eq:rhobga})). The behavior can be loosely divided into three distinct phases: A free expansion phase in which the effects of drag are minimal, a reshaping phase in which the lightest sections of the UDS begin to experience some drag, and a stalling phase in which the entire stream halts its radial motion and begins to orbit with the background flow.

During the free expansion phase, each individual segment continues along its original hyperbolic trajectory, leading to an extreme elongation of the stream in which its length to width ratio can exceed $10^{5}$, depending on when and where recombination occurs within the stream. It continues to span an angle $\theta_{\rm arc}$ during this period, and combined with the elongation this produces a mildly-curved arc that is difficult to distinguish from a straight line.

Once a segment sweeps up a mass comparable to its own, it begins to slow down due to drag forces (left columns of Figures \ref{fig:mstails} and \ref{fig:gianttails}). Because the lightest portions of the stream correspond to what was once the outer layers of the star, they have linear densities that are a factor $|\rho|/\rho_{\rm max} \sim 100$ (in the MS case) times less than the heaviest portions, where $|\rho|$ and $\rho_{\max}$ are the average and maximum densities of the originally disrupted star, respectively. In full disruptions, the lightest portions of the stream correspond to the most-unbound debris, whereas UDSs from partial disruptions feature low-density regions at both the most- and least-unbound ends of the stream due to the surviving stellar core \citep[see Figure 10 of][]{Guillochon:2013a}. This causes one (or both) ends of the UDS to slow down relative to the heavier midsection, deforming the original arc shape into a loop. At first, this reshaping results in a slightly increased effective drag force, this comes as a result of  the midsection segments rotating by almost an angle $\pi$ relative to their original orientation, which temporarily causes them to travel broadside relative to their motion within the background medium. Once the segments have reoriented to travel in a direction parallel to their length, the orientation component of the drag force per unit mass (Equation~(\ref{eq:length})) is reduced to close to its original value.

The UDS continues to travel outwards with a loop-like shape until even the heaviest portions of the stream begin to decelerate from the mass they sweep up (right columns of Figures \ref{fig:mstails} and \ref{fig:gianttails}). This distance varies greatly from event to event depending on the mass of the star that was disrupted, the fraction of mass the star lost at periapse, and the initial density of the star. For MS stars, the minimum travel distance is $\sim 1$~pc, with some UDSs extending all the way to $\sim 100$~pc before stalling. Giant stars show a similar range of terminal distances, although their outward motion terminates a factor of a few closer to the black hole owing to their greater sizes and generally lower fractional mass loss.

\begin{figure}
\centering\includegraphics[width=0.9\linewidth,clip=true]{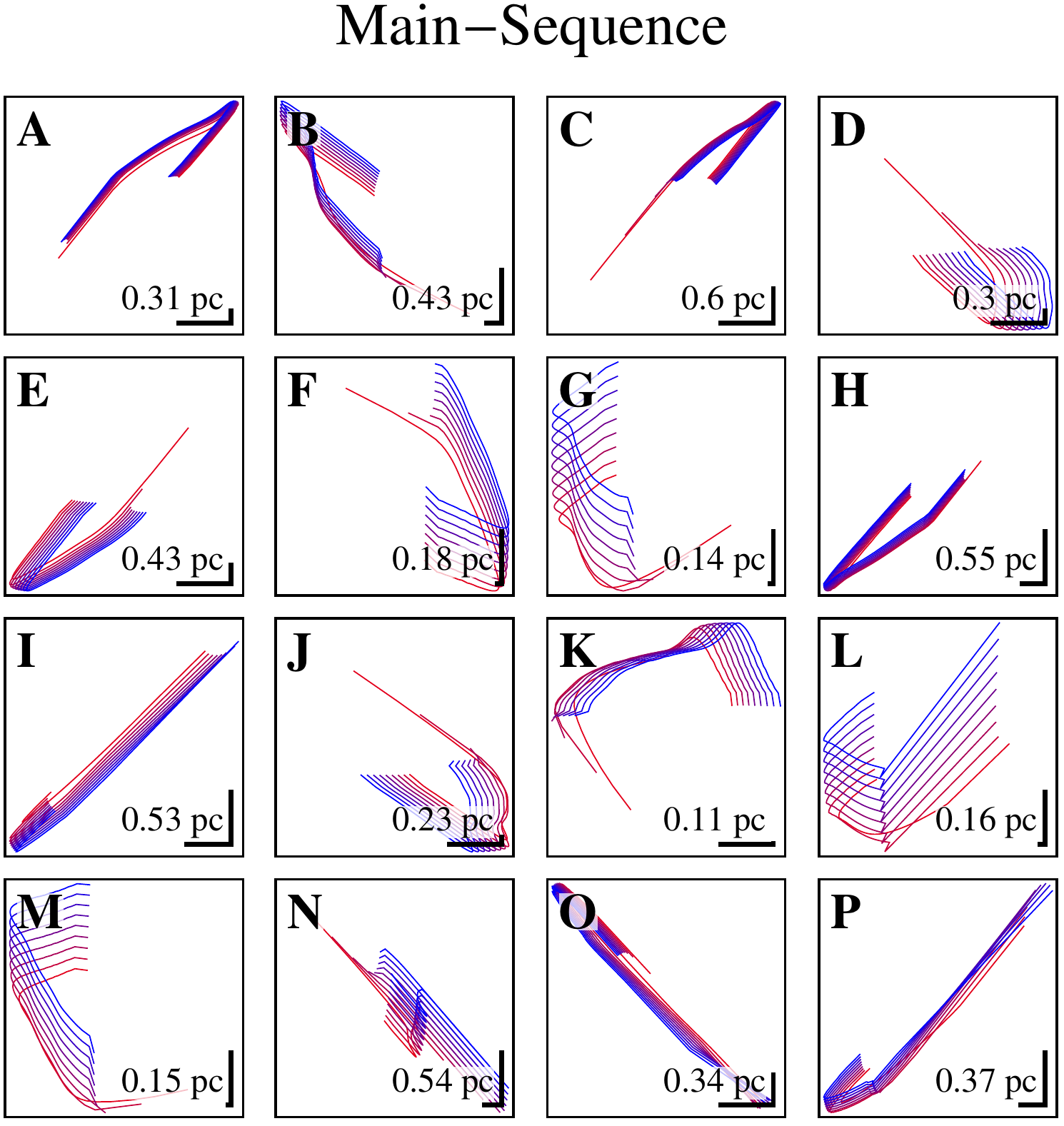}\\\vspace{1em}
\includegraphics[width=0.9\linewidth,clip=true]{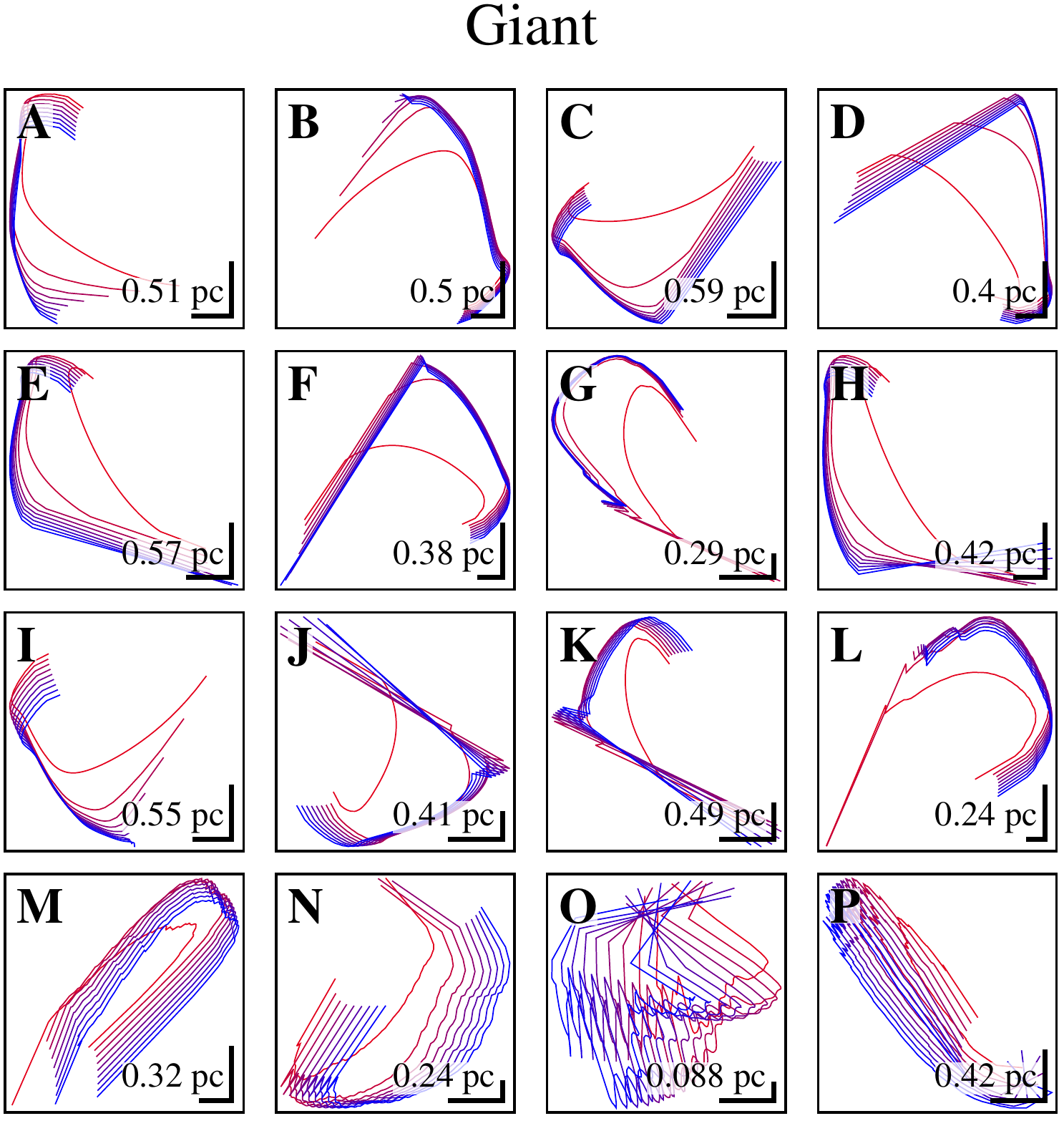}
\caption{Identical to the rightmost columns of Figures \ref{fig:mstails} and \ref{fig:gianttails}, but for a selection of UDSs generated by the disruption of MS and giant stars for our case~B background density, Equation~(\ref{eq:rhobgb}).}
\label{fig:walltails}
\end{figure}

We follow the evolution of UDSs until the outward radial motion of all individual segments ceases, i.e. $\max(\{{{\bf x}_{i}\,\cdot\,{\bf \dot{x}}_{i}},\;\ldots,\;{{\bf x}_{N}\,\cdot\,{\bf \dot{x}}_{N}}\}) < 0$, implying that the stream has deposited the entirety of its kinetic energy and momentum into the ISM. As a consequence, the total amount of energy and momentum deposited is identical for a given set of disruption parameters regardless of the background ISM density profile. But while the final energetics are similar, the morphology of the outgoing streams are notably different for streams that traverse the lower-density ISM (our case~A) and those that run into a region of high density (our case~B), as depicted in Figure~\ref{fig:walltails}. Whereas case~A often resulted in UDSs that were highly elongated in the radial direction, case~B UDSs are more compact once their outward motion has ceased. This arises because all of the individual stream segments halt at approximately the same distance, the location of the ``wall'' where the amount of mass they sweep up approaches their own mass very quickly. As we'll describe in Section \ref{sec:udr}, this difference in UDS morphology affects the evolution of the kinetic luminosity that ultimately powers any resulting remnants.

\begin{figure}
\centering\includegraphics[width=0.49\linewidth,clip=true]{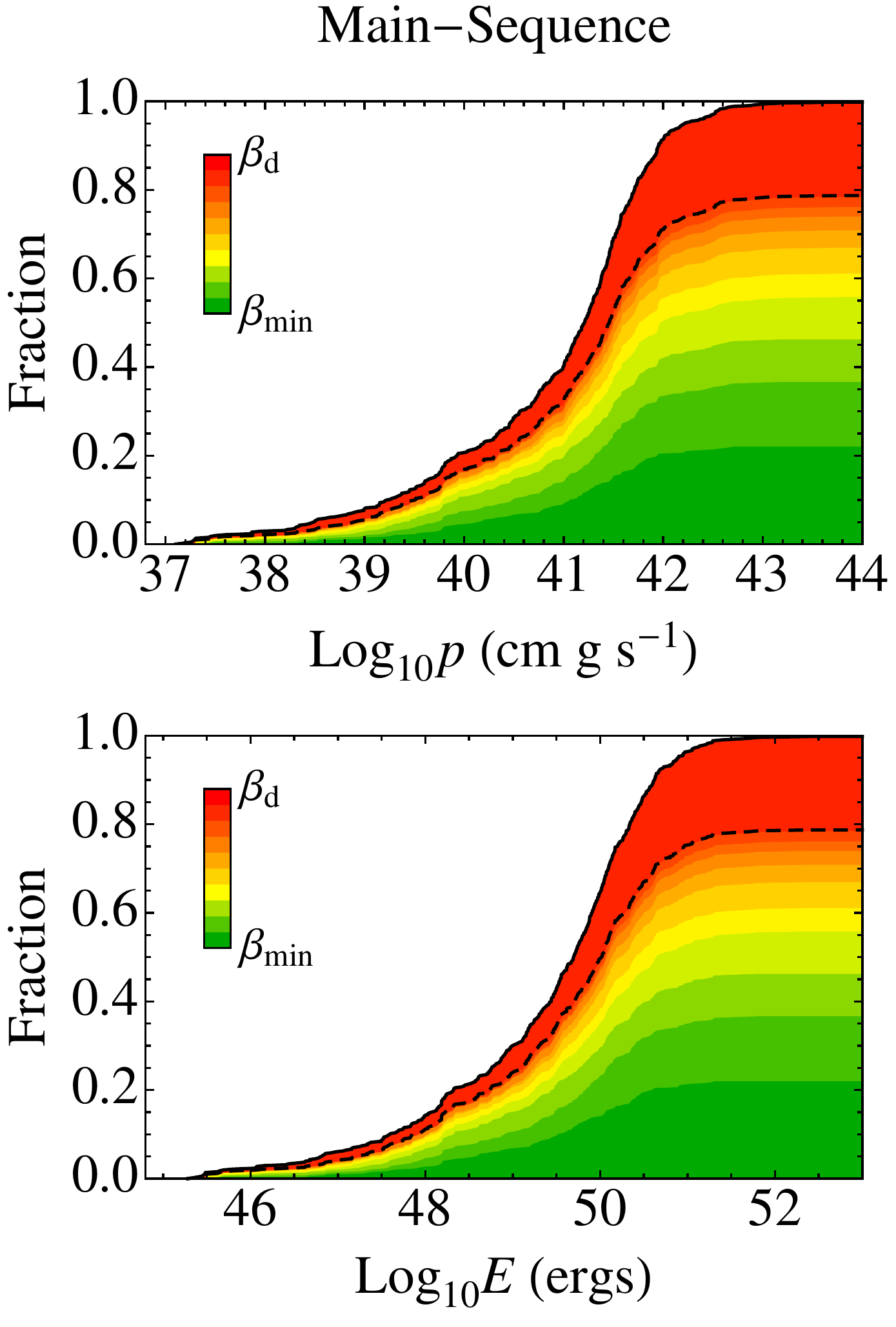}~~\includegraphics[width=0.49\linewidth,clip=true]{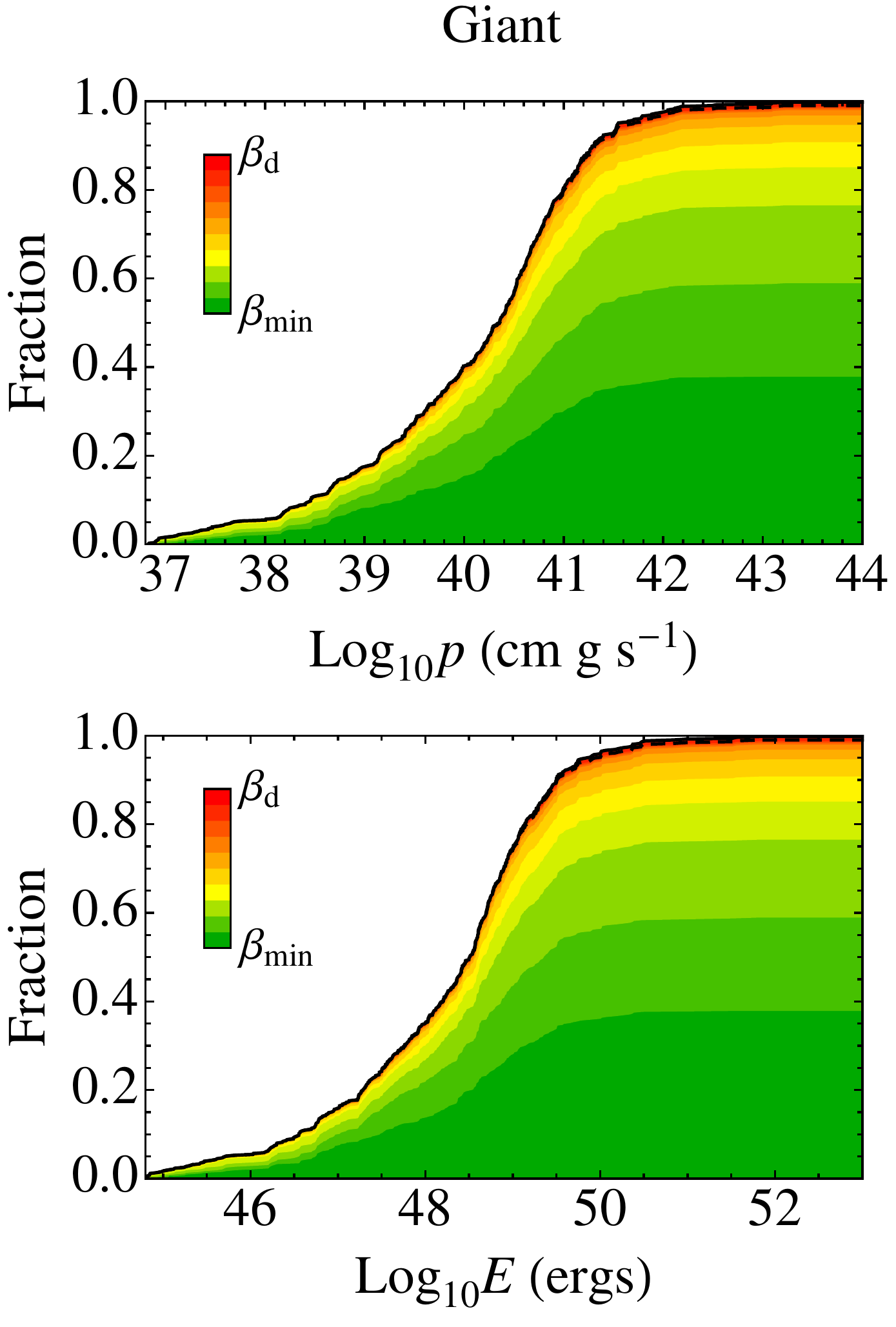}
\caption{Cumulative distribution functions of momentum $p$ (top panels) and energy $E$ (bottom panels) injected into the ISM by UDSs. The left panels show the momentum and energy injected by main-sequence star disruptions, whereas the right panels show these quantities for giant disruptions. The fraction of events associated with a given impact parameter is indicated by the colored shading of each distribution, with grazing encounters in which a tiny fraction of the star's mass is lost ($\beta = \beta_{\min}$) shown in green, and complete disruptions ($\beta > \beta_{\rm d}$) shown in red above the dashed line.}
\label{fig:ephist}
\end{figure}

The distribution of injected momenta and kinetic energy is shown in Figure~\ref{fig:ephist}. The median momentum and kinetic energy deposited by a MS (giant) disruption is found to be $2~\times~10^{41}$ ($2 \times 10^{40}$)~cm~g~s$^{-1}$ and $5~\times~10^{49}$ ($3 \times 10^{48}$)~erg, respectively. Many events are found to be sub-energetic, with 6\% (16\%) of MS (giant) disruptions yielding $< 10^{47}$~erg, most of these disruptions involve either low-mass stars or partial disruptions in which only a small fraction of the star's mass is ejected. Supernova-like kinetic energies of $10^{51}$~erg are realized for 4\% of MS disruptions, but almost never ($0.6$\%) for giant disruptions. Over-energetic disruptions are very rare, with less than a percent of MS disruptions, and no giant disruptions, yielding $10^{52}$~erg of energy. We find that $10^{53}$~erg, the fiducial value adopted by \citet{Khokhlov:1996a}, is never realized even for the deepest MS encounters of the most massive stars.

\begin{figure*}
\setlength{\tabcolsep}{0em}
\centering\begin{tabular}{>{\centering\arraybackslash}m{0.04\linewidth}>{\centering\arraybackslash}m{0.425\linewidth}>{\centering\arraybackslash}m{0.425\linewidth}}
& {\Large \bf \hspace{1.5em}Case~A} & {\Large \bf \hspace{1.5em}Case~B}\\
\rotatebox{90}{\Large \bf \hspace{1.5em}MS} & \includegraphics[width=\linewidth,clip=true]{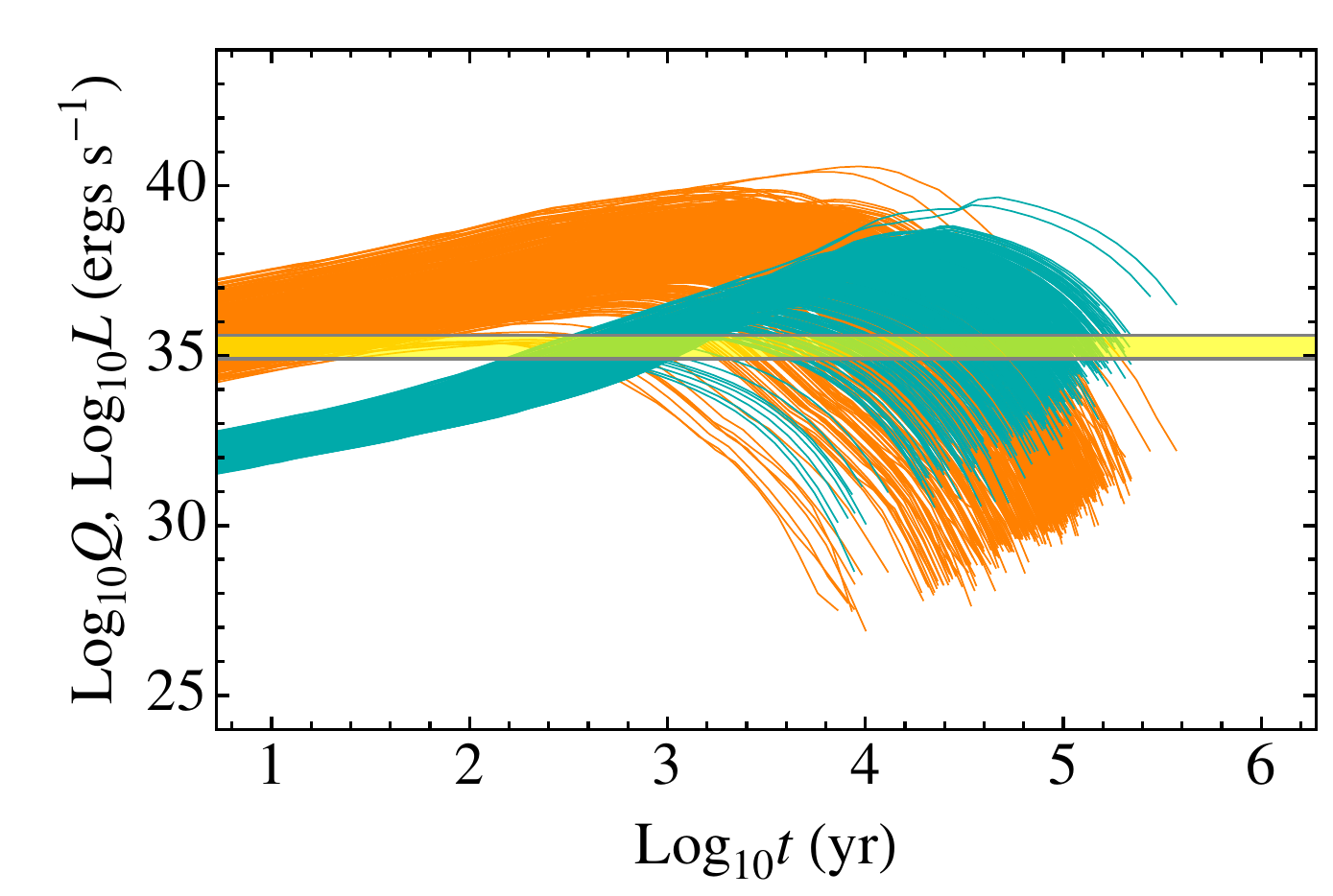} & \includegraphics[width=\linewidth,clip=true]{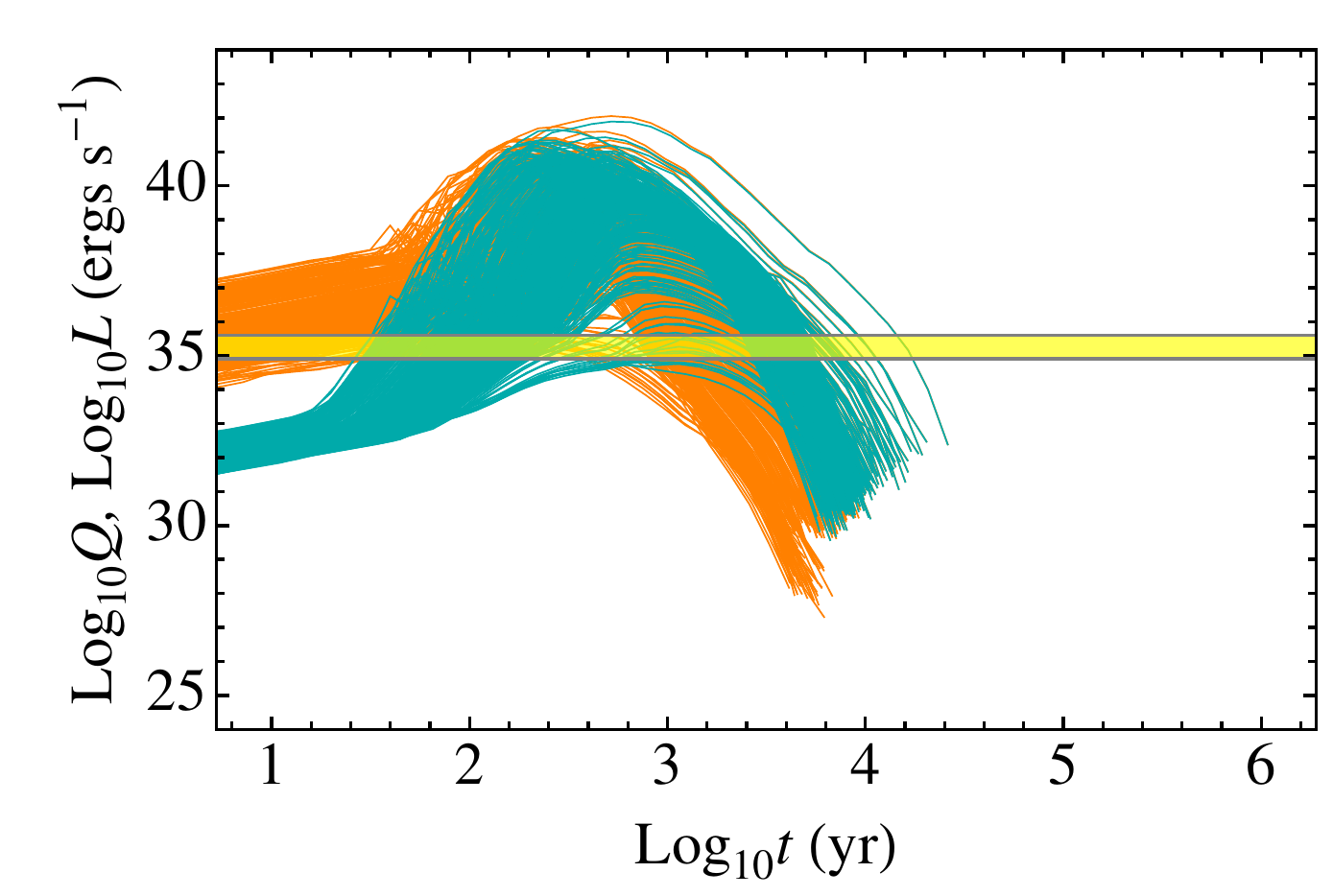}\\
\rotatebox{90}{\Large \bf \hspace{1.5em}Giant} & \includegraphics[width=\linewidth,clip=true]{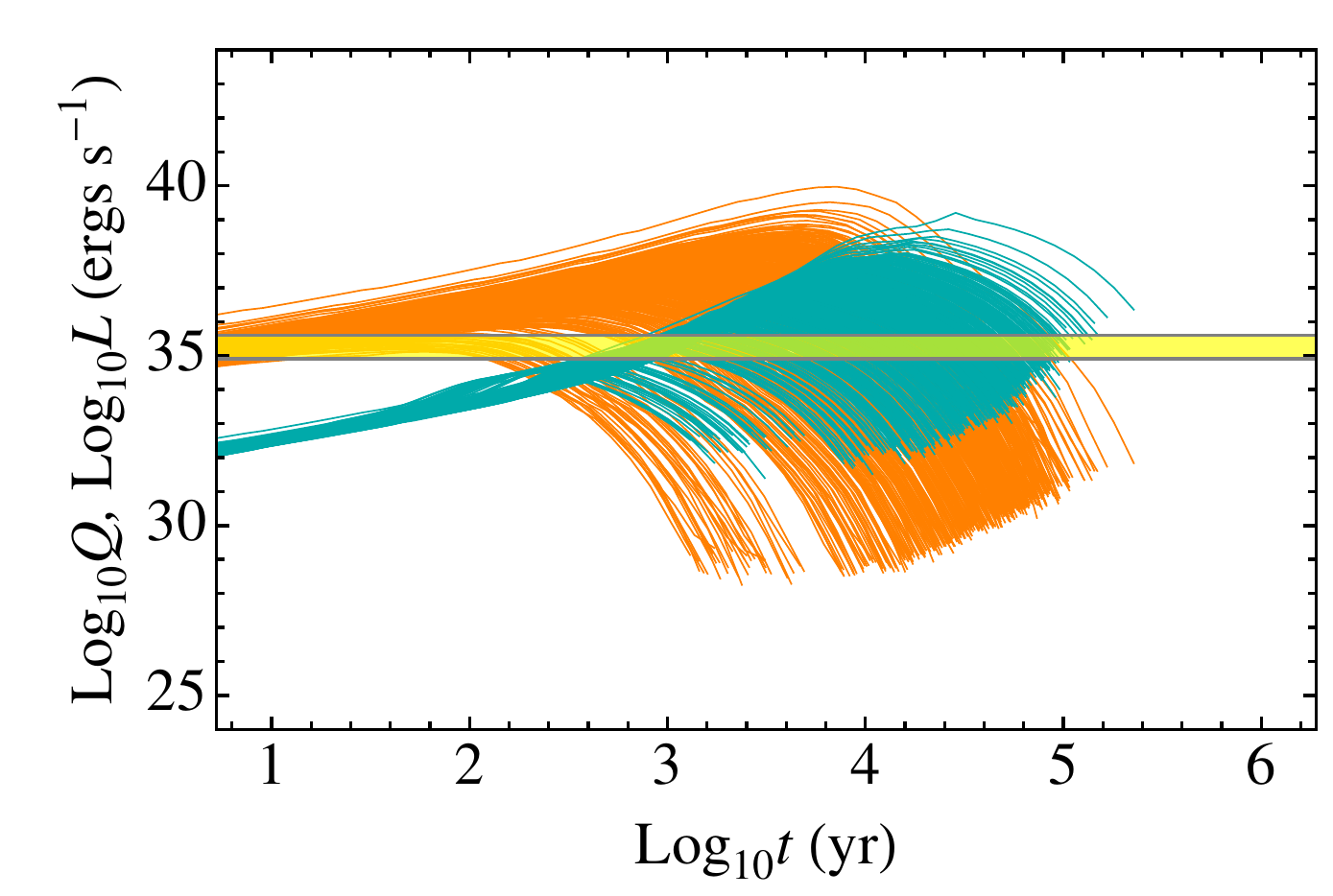} & \includegraphics[width=\linewidth,clip=true]{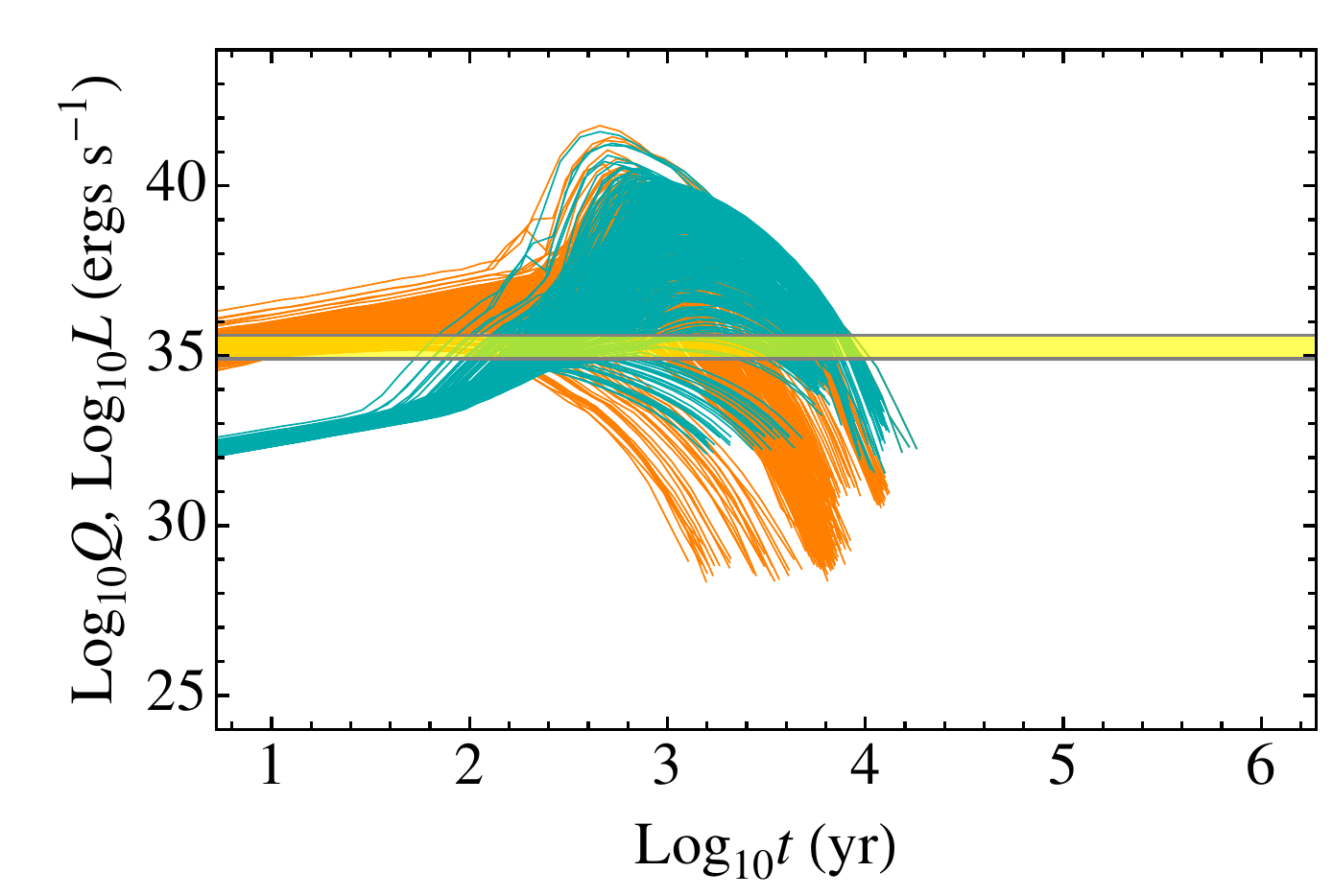}\\
\end{tabular}
\caption{Energy injected by each UDS compared to energy radiated as a function of time since disruption. Each panel shows every run from each of the described star/background density combinations, with the top row showing MS UDSs, the bottom row showing Giants, the left column the case~A density assumption, and the right column showing case~B. The orange curves in each panel show the heating rate $Q$ of the ambient medium for each UDS in the ensemble as determined by our Monte Carlo calculation, while the aqua curves show the cooling rate $L$ (Equation~(\ref{eq:lum})) of the resulting UDRs. A range of plausible values for \se's bolometric luminosity, with the lower limit taken from \citet{Maeda:2002a} and the upper limit taken from \citet{Fryer:2006a}, is shown by the yellow band in each panel.}
\label{fig:dc}
\end{figure*}

\section{Unbound Debris Remnants}\label{sec:udr}
As shown in Section \ref{sec:uds}, each tidal disruption injects a tremendous amount of energy into the ISM surrounding the black hole, with $10^{50}$~erg per event being typical, and some disruptions yielding as much as $10^{52}$~erg of energy. The typical stalling distance $r_{\rm stall}$ of MS UDS for case~A $\simeq 20$~pc, and as $\theta_{\rm arc} \simeq 0.1$ for $q = 4 \times 10^{6}$, this yields a cross-sectional width of $\theta_{\rm arc} r_{\rm stall} = 2$~pc. At this time, the stream's profile has only expanded by a factor $c_{\rm s,rec} t_{\rm stall}$, which is $\simeq 0.05$~pc, and hence the deposition region is typically several pc long, a few pc wide, and a tenth of pc tall, comparable to the size of a SNR as it transitions from the free-expansion phase to the Sedov-Taylor phase \citep{McKee:1995a}. What sort of ``unbound debris remnant'' (UDR) might we expect originate from the interaction of UDSs with their environments? To determine this, we consider two similar types of kinetic energy injection into the ISM: Supernovae and jets.

The blast waves produced by SNe result in SNRs that are only mildly asymmetric at late times even in the most extreme cases \citep{Lopez:2009a}, even though the blast waves themselves can be very asymmetrical at early times, especially if the supernova also produced a jet \citep{Maund:2009a,Lopez:2013a}. This simplicity arises from the fact that SNRs radiate only a small fraction of the total kinetic energy they possess until many thousands of years after the explosion. This inability to radiate means that the initial remnant, regardless of its original shape, has a very large internal sound speed. Gas in the initial remnant expands isotropically due to its high internal pressure, quickly reaching a spherical shape with a radius described by the Sedov-Taylor solution \citep{Sedov:1946a,Taylor:1950a,Ostriker:1988a},
\begin{equation}
R_{\rm Sedov-Taylor} = 1.15 \left(\frac{E}{\rho_{\rm bg}}\right)^{1/5} t^{2/5},\label{eq:rst}
\end{equation}
where the equation above is presented in CGS.

But even in the most asymmetrical supernovae, some fraction of the outgoing blast is radiated in every direction, whereas UDSs are decidedly unidirectional when they are first ejected. Jets differ significantly from supernova blast waves in that they are highly-collimated and traveling at very high Mach numbers ${\cal M}$ with velocities comparable to the escape velocities of the objects that produce them \citep{Livio:1999a}, and thus their dynamical evolution may provide a closer analogue to UDSs. Early in their evolution jets travel at velocities comparable to their initial velocity and do not resemble SNRs, instead producing a narrow cone-like structure with an opening angle equal to $1/{\cal M}$. However, when jets sweep up a mass comparable to their own, they slow down relative to the background and form an overpressure located at the jet's head, which can be continuously energized if the jet remains powered indefinitely, or until the powering time becomes comparable to the radiative cooling time \citep{Heinz:2014a}. If the jet has a finite lifetime that is shorter than the radiative time, this head region will eventually contain a large fraction of the jet's total energy and momentum. The overpressure in this region then acts as an energy-driven spherical explosion \citep{Avedisova:1972a,Castor:1975a}, with radius
\begin{equation}
R_{\rm head} = 0.76 \left(\frac{\dot{M} v^{2} t^{3}}{2 \rho_{\rm bg}}\right)^{1/5}\label{eq:rinit},
\end{equation}
assuming a constant density of the ISM $\rho_{\rm bg}$ and kinetic luminosity $\dot{M} v^{2}$. If the remnant evolution remains adiabatic while energy is injected into the overpressure region, Equation~(\ref{eq:rinit}) can be written in terms of a total kinetic energy $E = \dot{M} v^{2} t/2$,
\begin{equation}
R_{\rm head} = 0.76 \left(\frac{E}{\rho_{\rm bg}}\right)^{1/5} t^{2/5}.
\end{equation}
Comparison of the above with Equation~(\ref{eq:rst}) shows the similarity of the two solutions, with the only difference being the numeric constant. This similarity arises because of the fundamental assumption made in both cases that no energy is radiated before the bulk of the kinetic energy is deposited into the external medium, i.e. the evolution is adiabatic.



\subsection{A model for UDR}

As we have described, UDSs are somewhat in between the purely spherical explosions that produce SNRs and the highly-collimated flows of a jet. If the region that they heat via their passage is unable to cool on a timescale shorter than the time it takes to deposit the UDS's kinetic energy, the resulting remnant will become round and very much resemble SNRs or the bubbles found at the heads of stalled astrophysical jets. As shown in Equation \ref{eq:vmax}, the maximum velocity of a UDS is 8,000~km~s$^{-1}$, however disruption simulations show that the average kinetic energy is about a order of magnitude smaller than this value \citep{Guillochon:2013a}, resulting in a typical outgoing velocity of $\sim 2,\!500$~km~s$^{-1}$. This corresponds to a post-shock temperature (assuming a $\gamma = 5/3$, adiabatic fluid) of 
\begin{equation}
T = 2 \times 10^{8}~\left(\frac{v}{{\rm 2,\!500~km~s}^{-1}}\right)^{2}~K.\label{eq:t}
\end{equation}
At this temperature, metal-line cooling and Bremsstrahlung cooling are comparable to one another, with Bremsstrahlung being more important at solar metallicity \citep{Sutherland:1993a}. We adopt a simple cooling function to account for both processes,
\begin{equation}
\Lambda = 10^{-22.75} \begin{cases}
\left(\frac{T}{10^{7}}\right)^{-1}&T < 10^{7}\\
\left(\frac{T}{10^{7}}\right)^{1/2}&T > 10^{7}
\end{cases}~{\rm erg~cm}^{3}~{\rm s}^{-1},
\end{equation}
where $T$ is the post-shock temperature.

In the first phase of evolution (free expansion, see Section \ref{sec:uds}) the velocity is comparable to the initial value, resulting in $T$ being significantly larger than $10^{7}$~K and cooling being dominated by Bremsstrahlung. The material struck by the UDS will be imparted some momentum by the interaction, causing the background matter to be swept up into the expanding blast generated by the outgoing stream. Like jets, this will likely form a ``head'' at the tip of the stream within which most of UDS's kinetic energy will be deposited. And also like jets, UDS motion is highly supersonic relative to the background, with Mach numbers ${\cal M} \sim 100$ being typical once the UDS leaves the black hole's sphere of influence. Thermal energy is injected into this region near the head of the UDS as it travels outwards resulting in the formation of a UDR, the UDR expands into the ambient medium at the velocity given by the Sedov-Taylor solution, $v_{\rm UDR} = dR_{\rm UDR}(t)/dt$, where $R_{\rm UDR}(t)$ is determined via Equation~(\ref{eq:rst}) and $E$ is set to the accumulated energy injected into the ambient medium,
\begin{equation}
E_{\rm UDR} (t) = \int_{0}^{t} Q(t') dt',
\end{equation}
where the energy injection rate $Q(t)$ is determined numerically from our Monte Carlo (Figure~\ref{fig:dc}, orange curves). If we assume a near-constant background density $\rho_{\rm bg}$ as the UDS stalls, the UDR expansion velocity is
\begin{equation}
v_{\rm UDR} (t) = \frac{2 E_{\rm UDR} + Q t}{5 t^{3/5} \rho^{1/5} E_{\rm UDR}^{4/5}},\label{eq:vudr}
\end{equation}
which yields $v_{\rm UDR} \propto t^{2/5}$ at early times when $Q t \gg E_{\rm UDR}$ and the Sedov-Taylor scaling $v_{\rm UDR} \propto t^{-3/5}$ when $Q t \ll E_{\rm UDR}$.

Because the volume of the natal UDR $V_{\rm UDR} \propto R_{\rm UDR}^{3}$ is small at early times, the total cooling rate $\Lambda n^{2} V_{\rm UDR}$ is many orders of magnitude smaller than the heating rate for first few hundred years (Figure~\ref{fig:dc}, aqua curves). Once the UDS enters the stalling phase its velocity $v$ decreases from its initial value via drag, $Q$ decreases significantly, and eventually the expansion velocity of the UDR $v_{\rm UDR}$ exceeds the outgoing UDS velocity $v_{\rm UDS}$, which is the velocity that now determines the post-shock temperature. The UDRs quickly transitions to a phase where it is cooled primarily by atomic lines as the expansion velocity drops $\propto t^{-3/5}$ (Equation~(\ref{eq:vudr})). Once line-cooling dominates, the timescale when the remnant becomes radiative $\tau_{\rm rad}$ can be determined by solving $t = E(t)/(L(t) - Q(t))$ for $t$. We assume that once $t \gg \tau_{\rm rad}$ that the remnant quickly assumes a steady state where $L = Q$. This motivates the following functional form for the cooling rate,
\begin{equation}
L = \Lambda n^{2} V_{\rm UDR} \exp \left[-t/\tau_{\rm rad}\right] + Q \left[-\tau_{\rm rad}/t\right],\label{eq:lum}
\end{equation}
where the first term dominates when $t \ll \tau_{\rm rad}$ and the second term dominates when $t \gg \tau_{\rm rad}$. Once a UDR becomes radiative ($t > \tau_{\rm rad}$), we assume that it transitions into a pressure-driven phase \citep{McKee:1977a} where
\begin{equation}
R_{\rm UDR} = R_{\rm Sedov-Taylor} (\tau_{\rm rad}) (t/\tau_{\rm rad})^{2/7},\label{eq:rrad}
\end{equation}
with $R_{\rm Sedov-Taylor} (\tau_{\rm rad})$ being calculated using Equation~(\ref{eq:rst}).

For a simple Sedov-Taylor solution, the time at which the structure would become radiative is \citep{Blondin:1998a}
\begin{equation}
\tau_{\rm rad} = 1.7 \times 10^{4} \left(\frac{E}{10^{50}~{\rm erg}}\right)^{4/17} \left(\frac{n_{\rm bg}}{{\rm cm}^{-3}}\right)^{-9/17}~{\rm yr},\label{eq:radtime}
\end{equation}
where $n_{\rm bg} = \rho_{\rm bg}/\mu m_{\rm p}$ and we have scaled the fiducial energy to $10^{50}$~erg, as is appropriate for UDRs. Inspection of the peak luminosities in Figure~\ref{fig:dc} shows this timescale is very typical of UDRs, and demonstrates that the Sedov-Taylor solution is a good approximation to UDRs before they become radiative.

The timescale over which kinetic energy is injected into the ISM can be quite long for case~A for both MS stars and giants (Figure~\ref{fig:dc}, first column), with energy being deposited over a significant fraction of the UDS's final length. Individual pieces of the UDS may become radiative before others, especially if the UDS covers a broad angle and interacts with spatially-disconnected regions of enhanced density in the ISM, as might be expected in the CND. For UDSs that remain compact as they begin to slow down, most of the energy will be deposited into the same small region; MS UDSs are generally very streamlined and thus likely to behave this way (Figure~\ref{fig:mstails}). Additionally, despite the large time range over which the energy is deposited, the resulting remnant remains adiabatic even significantly after the time that $Q$ reaches its maximum, with $\tau_{\rm rad}$ typically being a factor of ten times longer than this time.

For case~B, the UDSs travel unimpeded until they encounter the wall, at which point they deposit most of their energy over a short distance and time, as shown in the right column of Figure~\ref{fig:dc}. In this case, both the timescale over which energy is injected and the timescale over which this energy is radiated are shorter owing to the larger ambient density. The resulting UDRs are also significantly smaller, with radius $\sim 10$ times smaller than case~A UDRs (and $10^{3}$ times less volume). In contrast with case~A, UDRs formed in case~B have radiative timescales that are comparable to the time at which $Q$ is maximized, this suggests that significant energy is still being deposited into the ISM even when case~B UDRs are at their brightest. But while these UDRs are brighter at peak, the time that they are observable is possibly only a few thousand years, making their detection in our own GC less likely unless the tidal disruption rate were enhanced.


\section{Impact on Galactic Nuclei}\label{sec:impact}
UDRs affect their host galaxies in much the same way that SNRs do, with the differences in their energy ($\sim 10\%$ of the energy per event, Figure~\ref{fig:ephist}), their rates \citep[$\sim 1\%$ the total MW SNe rate,][]{Stone:2014a}, and their location (concentrated in the GC). SNRs, especially those resulting from core-collapse SNe, tend to be located in regions of ongoing star formation \citep{Anderson:2015a}. The centers of galaxies are often actively star-forming, as is the case for our own GC, but star formation tends to be spread over a significant fraction of a star-forming galaxy's volume; by contrast we show that UDRs are always contained within $\sim 100$~pc of the galactic nucleus. For galaxies with no ongoing star formation, UDRs will outnumber SNRs, especially within the central regions. Because of the similarity of SNRs and UDRs, and given SNRs are important for accelerating cosmic rays, being the source of the far infrared-radio correlation, and driving turbulence in star-forming regions, the impact of UDRs in galactic centers (including our own) should be seriously considered.
\subsection{\se: SNR or UDR?}
\citet{Khokhlov:1996a} first proposed that \se might have originated from the interaction of the unbound debris of a tidally disrupted star with the surrounding medium. They presumed that the unbound debris possessed an energy equal to the energy spread at periapse even for deeply-penetrating encounters where $r_{\rm p} \sim R_{\ast}$, resulting in extremely large kinetic energies (up to $10^{53}$~erg). However, it was shown in \citet{Guillochon:2013a} and \citet{Stone:2013a} that the spread in binding energy is maximal for events where $\beta \simeq \beta_{\rm d}$, the impact parameter where the star is fully destroyed, with no increase in energy for events with $\beta > \beta_{\rm d}$. This reduces the amount of kinetic energy by a factor of $q^{1/3}$ relative to the assumption of \citet{Khokhlov:1996a}, which results in a typical energy of $\sim 10^{50}$~erg for \sa, as is found in our numerical results (Section \ref{sec:uds}). Additionally, \citet{Khokhlov:1996a} presumed that a solar mass star was fully disrupted, whereas most disruption events involve less-massive stars closer to the peak in the IMF ($M_{\ast} \sim 0.3 M_{\odot}$) and a $\beta$ where only a fraction of the star's mass is ejected.

\citet{Mezger:1989a} suggested that the energy of the blast that produced \se would need to be $\sim 10^{52}$~erg if the surrounding dust shell (with density $\sim 10^{4}$~cm$^{-3}$) was evacuated by the blast. But X-ray observations of \se have shown that the total energy contained within the region is no greater than $10^{50}$~erg \citep{Maeda:2002a,Sakano:2004a}, a value very similar to the median energy deposited by a single UDS (Figure~\ref{fig:ephist}).

\begin{figure}
\centering\includegraphics[width=0.9\linewidth,clip=true]{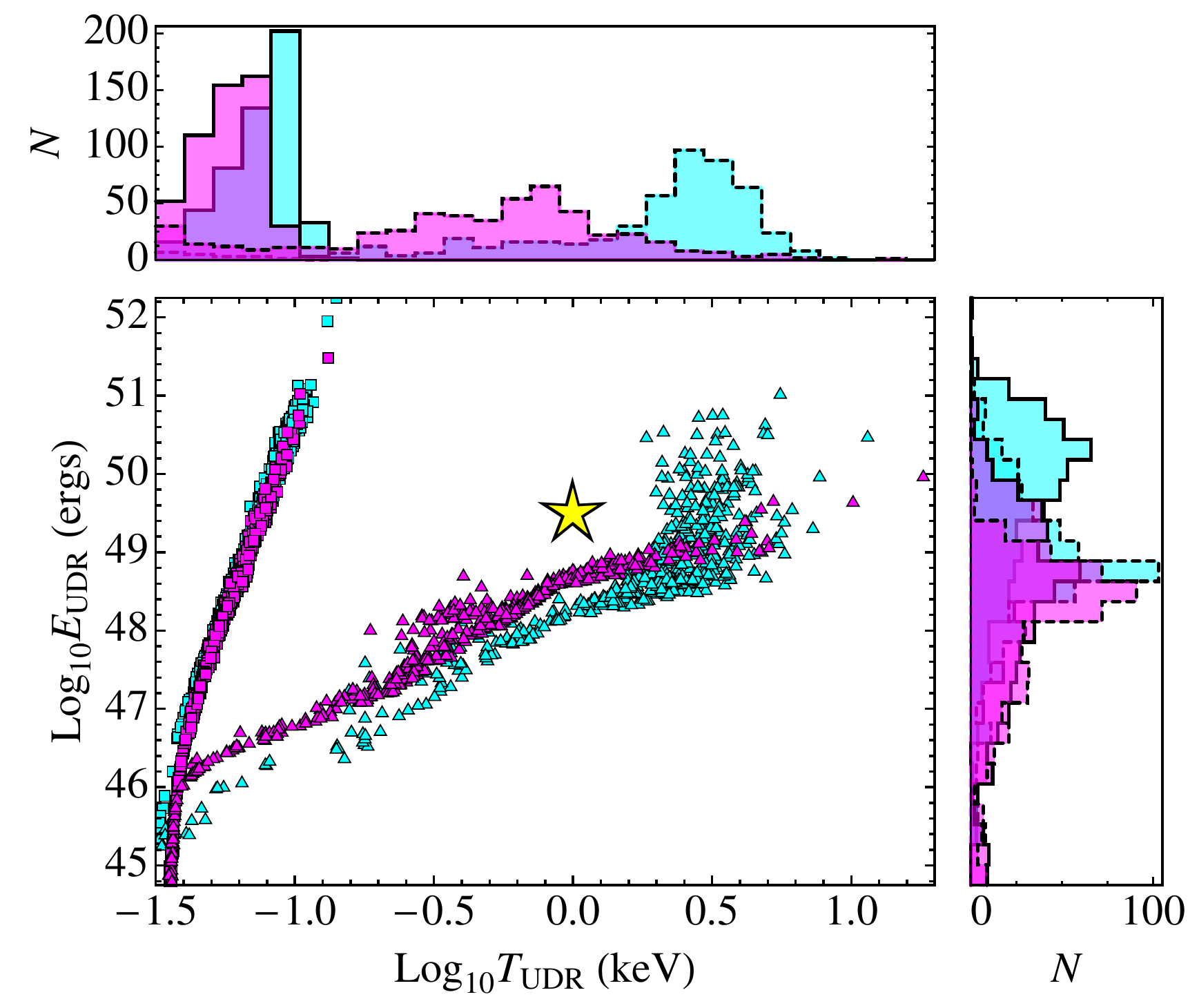}\\
\centering\includegraphics[width=0.9\linewidth,clip=true]{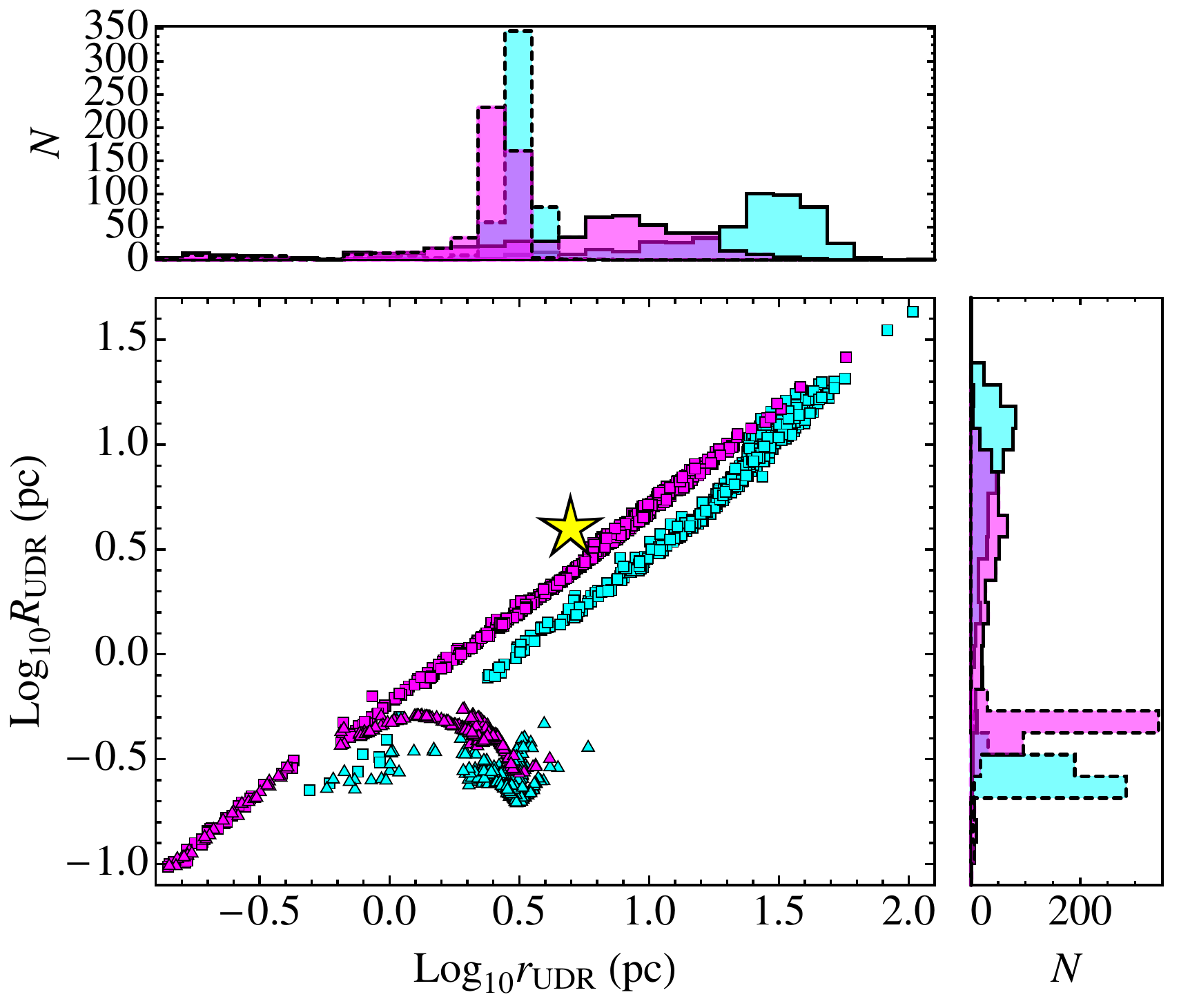}
\caption{Fundamental properties of UDRs resulting from tidal disruptions. The two panels show four quantities calculated at the time the UDRs becomes radiative (peak of $L$ curves in Figure~\ref{fig:dc}), with cyan (magenta) corresponding to MS (giant) UDR, and the squares (triangles) corresponding to case~A (B). The histograms in each panel bin the points along the $x$ and $y$ axes, with the solid histograms corresponding to case~A and the dashed histograms corresponding to case~B. The quantities shown are the thermal energy content $E_{\rm UDR}$, temperature $T_{\rm UDR}$, UDR size $R_{\rm UDR}$, and distance of the UDR center from \sa, $r_{\rm UDR}$. \se, with values taken from \citet{Maeda:2002a}, \citet{Park:2005a}, and \citet{Yusef-Zadeh:1999a}, is shown by a yellow star in each panel.}
\label{fig:comp}
\end{figure}

Figure~\ref{fig:comp} shows the energy content $E_{\rm UDR}$, temperature $T_{\rm UDR}$, size $R_{\rm UDR}$, and distance $r_{\rm UDR}$ from the GC of UDRs at the time they become radiative as compared to \se, which is represented by the yellow star in both panels. As shown in Figure~\ref{fig:dc}, \se is likely to be in a phase just prior to, or just after, the time it becomes radiative depending on its true age. The temperature of \se, $\sim$~1~keV \citep{Park:2005a}, lies between our case~A and case~B scenarios, with case~A predicting temperatures of 100~eV and case~B predicting temperatures of 3~keV given \se's present-day energy content. If \se is pre-radiative, then the velocity of the outgoing shock will be faster than at $t = \tau_{\rm rad}$, resulting in higher temperatures, whereas \se being post-radiative would suggest low present-day temperatures. The fact that temperature of \se is bracketed by our two cases suggests that perhaps the stream that formed the remnant had encountered a density intermediate to the two cases. \se's distance from the GC is estimated to be $\sim 5$~pc \citep{Yusef-Zadeh:1999a}, this distance is in agreement with case~A giant UDSs but in tension with case~A MS UDSs that typically reach tens of pc before stalling. For both MS and giant UDSs, case~B typically yields distances comparable to the distance we place our wall condition, which is slightly closer to the black hole than \se's observed distance and was motivated by the location of the CND. Given the better agreement with the case~B temperature and stalling distance, we reason that the UDS that would have produced \se likely interacted with at least some dense gas on its way out. There is significant evidence that \se has in fact interacted with dense molecular gas in the GC, such as the masing gas at its edges which is typically associated with such interactions \citep{Yusef-Zadeh:1999a}.

Upper limits in the iron abundance of \se have also been determined \citep{Park:2005a,Koyama:2007a}, suggesting that $< 0.15 M_{\odot}$ of iron lies within the iron-enhanced core region, and $< 0.27 M_{\odot}$ of iron is contained within the full remnant. However, the $0.15 M_{\odot}$ upper limit was obtained by assuming that the core of \se was {\it entirely iron} \citep{Park:2005a}, whereas a $\sim 5$ times enhancement in iron relative to solar (as indicated by spectral fitting) and a core mass of a few solar masses suggests a more pedestrian $M_{\rm Fe} \sim 0.01 M_{\odot}$. The full remnant, which includes perhaps a few tens of solar masses of material swept up from the ISM, has only a mildly enhanced iron abundance ($\sim 2$~--~3), as might be expected for the heavily-recycled gas in the GC, although evidence for such an enhancement is still tentative \citep{Genzel:2010b}. While a recent study by \citet{Do:2015a} found a number of metal-poor stars at the GC, the majority of stars have at least a solar metallicity, with the median value being super-solar ([M/H] = 0.4), with some stars being up to ten times as metal-rich as the Sun, $[{\rm M/H}] \simeq 1$, although these higher-metallicity measurements come with considerable systematic uncertainty. Previous measurements of metallicities in the GC have found more modest enhancements, [{\rm M/H}] = 0.14 \citep{Cunha:2007a,Ryde:2014a}.

If we use the quoted values for the total energy, ejecta mass, and current constraints on the iron content of \se, the remnant is consistent with a UDR that was produced by the full disruption of a $[{\rm M/H}] \sim 0.7$, $M_{\ast} \sim 3 M_{\odot}$ star that launched a UDS which interacted to some degree with the CND. These parameters do not uniquely classify \se as a UDR; indeed the currently favored interpretation of a core-collapse SNR is equally capable of explaining the remnant. But because the UDR scenario is able to reproduce all of \se's salient features, it is reasonable to perform collect additional observational data to settle \se's origin.

We propose three possible ways to test whether \se is a SNR or UDR: first, there's the ``cannonball,'' a compact radio and non-thermal X-ray source that appears to be emerging from \se's geometric center \citep{Park:2005a,Nynka:2013a}, which has been interpreted as a runaway pulsar that would be expected to be produced by a CC SNe. However, the inferred three-dimensional velocity of the cannonball is poorly constrained by weak constraints on \se's age. The cannonball's observed proper motion is 500~km~s$^{-1}$ \citep{Zhao:2013a} with a travel direction compatible with \se's center, but its radial velocity component is not known, and indeed the magnitude of its three-dimensional velocity vector could be significantly greater, perhaps as large as $10^{3}$~km~s$^{-1}$, a velocity that is very rarely achieved in simulations of CC SNe \citep{Nordhaus:2012a,Wongwathanarat:2013a}. Conversely, UDS velocities are naturally in this range, a few $10^{3}$~km~s$^{-1}$, with some parts of the UDS still being in motion even after the bulk of the kinetic energy has been deposited into the ambient medium (after the peak in $Q$, Figure~\ref{fig:dc}). In particular, the end of the loop-like structure of the UDS, which streamlines significantly as it travels outwards (e.g. Figure~\ref{fig:mstails}), could potentially produce a cannonball-like feature. Of course, it is entirely possible that the cannonball is not associated with \se at all, although its proximity and the morphology of the surrounding gas does suggest a connection. A second way to distinguish the two possibilities is via a more-complete measurement of its composition. While a metallicity a few times solar is plausible in the GC where metal-rich stars are common, a tenth of a solar mass of iron would definitively eliminate the UDR possibility. And while most radioactive products resulting from a SNe will have decayed given \se's advanced state, $^{59}$Ni may still be detectable even $\sim10^{5}$~yr after the explosion \citep{Fryer:2006a}. Lastly, hydrodynamical simulations of the UDR scenario in our GC \citep[similar to those performed for the SNR scenario, see e.g.][]{Plewa:2002a,Rockefeller:2005a,Fryer:2006a,Rimoldi:2015a} could yield valuable information on the expected morphology of the UDS and its resulting UDR, and could test whether a feature like the cannonball could be produced from the highest-velocity UDS material.

\subsection{Particle acceleration}
The collision of UDSs with either the low-density ISM (our case~A) or the high-density molecular clouds (case~B) will produce strong shocks, since the collisional velocity ($\sim0.03c$) greatly exceeds the sound speed in the ISM or molecular clouds. As the UDR expands into the ISM, the outgoing shell compresses the ambient gas, amplifying the magnetic field and driving turbulence. As in SNRs, this compression accelerates electrons and atomic nuclei via a Fermi process, producing a non-thermal distribution of particles with an extended power-law tail \citep{Bell:1978a,Spitkovsky:2008a}. High-energy particles whose Larmour radius is larger than the shock width will emerge as cosmic rays. Low-energy particles that are incapable of escaping the remnant remain confined to the UDR where they lose their energy primarily through synchrotron cooling of the electrons, with radiation extending from radio to X-ray frequencies. In the next two subsections we briefly summarize the expected contribution that UDRs will make to the budget of cosmic rays and the radio emission of their host galaxies.

\subsubsection{Cosmic ray production}
As about $10\%$ of the shock energy will be tapped to accelerate cosmic rays \citep{Hinton:2009a,Treumann:2009a}, a total amount of $10^{49}$ erg in cosmic rays will be produced after each TDE, assuming an average MS UDS kinetic energy of $10^{50}$ erg (see Figure~\ref{fig:ephist}). The propagation of such cosmic rays in the GC and the subsequent observational signatures have been studied in detail in a series of earlier works \citep{Cheng:2006a,Cheng:2007a,Cheng:2011a,Cheng:2012a}. In these works, a smooth ISM was assumed, which corresponds to our case~A, and it is predicted that the shocks are able to accelerate cosmic rays to PeV energies. In this case, the cooling of the cosmic rays, due to the inelastic collisions with the non-relativistic protons in the ISM, happens on a timescale of about $5\times10^7~(\rho_{\rm bg,A}/1.3\times10^{-24}~{\rm g~cm^{-3}})$ yr \citep{Cheng:2007a}. This cooling timescale is much longer than the timescale of cosmic-ray production, which is about $10^2-10^4$ years, as can be seen in our Figure~\ref{fig:dc}.  As a result, the cosmic rays are able to propagate to a large distance, ${\cal O}(1)$ kpc, from the GC \citep{Cheng:2006a,Cheng:2007a}.

As opposed to our case~A UDS which deposit cosmic rays in much the same manner as SNRs, a UDS interacting with a dense cloud (our case~B) behaves somewhat differently as the energy is delivered much more impulsively. The increased density in these clouds means that the radiative timescale for such remnants is significantly shorter (Equation~(\ref{eq:radtime}) and Figure~\ref{fig:dc}, right column), and as a result the production and propagation of cosmic rays is very likely confined to the molecular clouds they impact. As the proton densities inside the molecular clouds can be orders of magnitude higher than that in the ISM, so too the synchrotron cooling timescale is reduced by the same orders of magnitude. In a companion paper (Chen et al. in preparation) we will study this case in depth.

It has been pointed out that the cosmic rays propagating to kpc scale could potentially produce a structure mimicking the Fermi bubbles \citep[][]{Su:2010a}, with tidal disruptions potentially playing an important role \citep{Cheng:2011a}. We can derive the same conclusion based on the energetics of the UDSs in our simulations. If a TDE deposits on average $10^{49}$ erg of cosmic rays into the ISM, about $10^{53}~{\rm erg}$ of cosmic rays (i.e. $10^4$ UDSs) will have been injected into the ISM if we adopt a standard TDE rate of $2\times10^{-4}~{\rm yr^{-1}}$ for the GC \citep{Merritt:2010c} and a cooling timescale of $5\times10^7$ yr. This is $\sim$10\% the energy required to form the Fermi bubbles \citep{Su:2010a}, and implies a cooling rate of about $6\times10^{37}~{\rm erg~s^{-1}}$. Suppose $10\%$ of the cooling is due to $\pi^0$ production and the subsequent $\gamma$-ray radiation, where the $10\%$ is a standard efficiency for the cosmic rays in the energy band of $1-100$ GeV \citep{Kafexhiu:2014a}, this results in a $\gamma$-ray luminosity of $6\times 10^{36}~{\rm erg~s^{-1}}$, roughly 10\% the luminosity estimated for the Fermi bubbles in $\gamma$-ray bands \citep{Su:2010a}. While not dominant, these values suggest that UDSs do contribute significantly to the Fermi bubble energy budget.

It should be noted that TDE jets \citep[e.g.][]{Bloom:2011a,Brown:2015a,Pasham:2015a}, which are possibly a hundred times less common than the UDSs that are produced with every tidal disruption \citep{Cenko:2012b,Guillochon:2015b}, can inject similar amounts of cosmic rays in total, as they convert a significant fraction of the accreted star's rest mass into outgoing kinetic energy \citep{Farrar:2014a}. If the typical jetted TDE deposits $0.1 M_{\odot}$ of material onto the black hole and 1\% of the rest mass is converted to cosmic rays (assuming 10\% the rest mass is converted into kinetic energy and 10\% of that is converted into cosmic rays), then approximately $10^{53}$~erg of cosmic rays will be produced by these jets, equal to what is produced by UDSs. However, these cosmic rays will be deposited at a greater distance from the black hole as the outgoing TDE jets will be more-highly beamed and possess greater initial kinetic energies per event \citep[with Lorentz factor $\Gamma \sim 10$,][]{Burrows:2011b}.

\subsubsection{Radio emission from UDR}
Radio emission from supernovae is thought to arise at two different times. The interaction of a supernova's fast-moving ejecta with nearby dense gas can result in ``prompt'' emission, so-called radio supernovae \citep{Chevalier:1998a,Weiler:2002a}, hundreds of days after the explosion. Once the remnant has entered the Sedov phase of its evolution, synchrotron cooling of electrons accelerated in the outgoing blast wave provides longer-lasting radio emission that peaks typically hundreds of years after the explosion. Unlike core-collapse SNe, UDRs lack both the relativistic component (with $v_{\max} \sim 10^{4}$~km~s$^{-1}$, Equation \ref{eq:vmax}) and expand into a region that is not primed by a massive wind prior to the event; instead, the external density is set by the pre-existing ambient matter distribution surrounding the black hole. In our case~A, the density at the distance at which most UDSs stall is comparable to the local ISM density of 1~cm$^{-3}$, which is also similar to the environments inhabited by SNe Ia which are devoid of gas and which have not been observed to produce any prompt radio or X-ray signatures \citep{Reynolds:2008a}. And whereas SNe typically exit the free expansion phase after a few centuries, UDSs typically take $10^{3}$~yr to enter this phase, and therefore any prompt signatures would be spread over a period that is $\sim 10$ times longer than the corresponding free-expansion phase of SNR evolution. As a result, no significant prompt emission is likely to result from UDRs for dormant black holes. For black holes that are already accreting at rates closer to Eddington, the density of the ISM immediately surrounding the black hole may be significantly enhanced by the infalling gas and a prompt radio signature might be possible, although identifying this emission may be difficult given that an accreting black hole is likely already a strong source of radio.

Much of the radio emission from UDRs is likely to be produced by synchrotron cooling of relativistic electrons once the UDS has deposited the majority of its energy into the ISM 100 -- $10^{4}$~yr after the disruption. Once that occurs, the Sedov-Taylor solution provides a reasonable approximation to a UDR's properties, as described in Section \ref{sec:udr}. For SNRs, the exact fraction of kinetic energy that ends up heating electrons that eventually cool via synchrotron is likely dependent upon a number of parameters, including the mass and energy of the ejecta, density of the ISM, etc., which are expected to vary greatly from event to event. But despite these complications, there exists a simple empirical relationship between radio luminosity and SNR size, as originally characterized by \citet{Clark:1976a}. Because UDRs likely resemble SNRs once they are in their Sedov-Taylor phase, we use an updated version of this empirical relationship \citep[Equation~(2)]{Pavlovic:2014a} to estimate the radio luminosity of UDR,
\begin{align}
\Sigma_{\nu} =~&7 \times 10^{-11} \left(\frac{R_{\rm UDR}}{\rm pc}\right)^{\beta}\nonumber\\
&\left(\frac{\nu}{\rm 1~GHz}\right)^{\alpha} {\rm erg~cm}^{2}~{\rm Hz}^{-1}~{\rm Sr}^{-1},
\end{align}
where $\Sigma_{\nu}$ is the surface density at a given frequency $\nu$, $\beta = -5.2$ is the $\Sigma$--$D$ slope found by \citet{Pavlovic:2014a}, and we have set the spectral index $\alpha = -0.5$. The radio luminosity $L$ of each UDR is then
\begin{align}
L &= 4 \pi^{2} R_{\rm UDR}^{2} \nu \Sigma_{\nu}\\
&= 200 L_{\odot} \left(\frac{\nu}{\rm GHz}\right)^{0.5}\left(\frac{R_{\rm UDR}}{\rm pc}\right)^{-3.2}.\label{eq:lnu}
\end{align}
This luminosity cannot of course exceed the injection of energy into accelerated particles, so we ceiling $L$ to values no greater than 10\% the maximum energy injection rate $Q_{\rm max}$ (Figure~\ref{fig:dc}), this most affects the luminosity shortly after the Sedov-Taylor time when the remnant is small and Equation~(\ref{eq:lnu}) diverges.

\begin{figure}[t]
\centering{\Large \bf \hspace{1.5em}MS, Case~A}\\
\centering\includegraphics[width=\linewidth,clip=true]{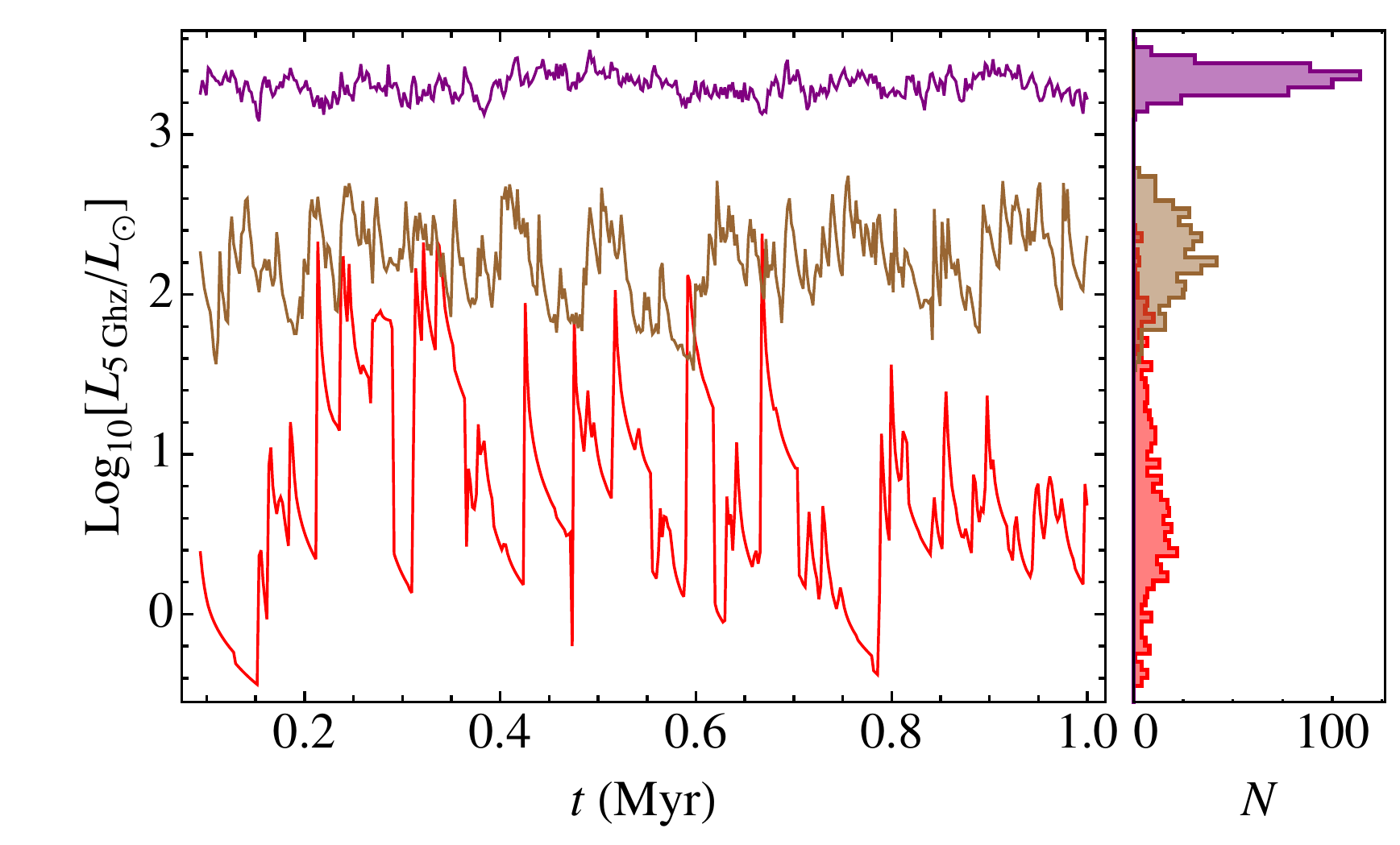}\\
\centering{\Large \bf \hspace{1.5em}MS, Case~B}\\
\centering\includegraphics[width=\linewidth,clip=true]{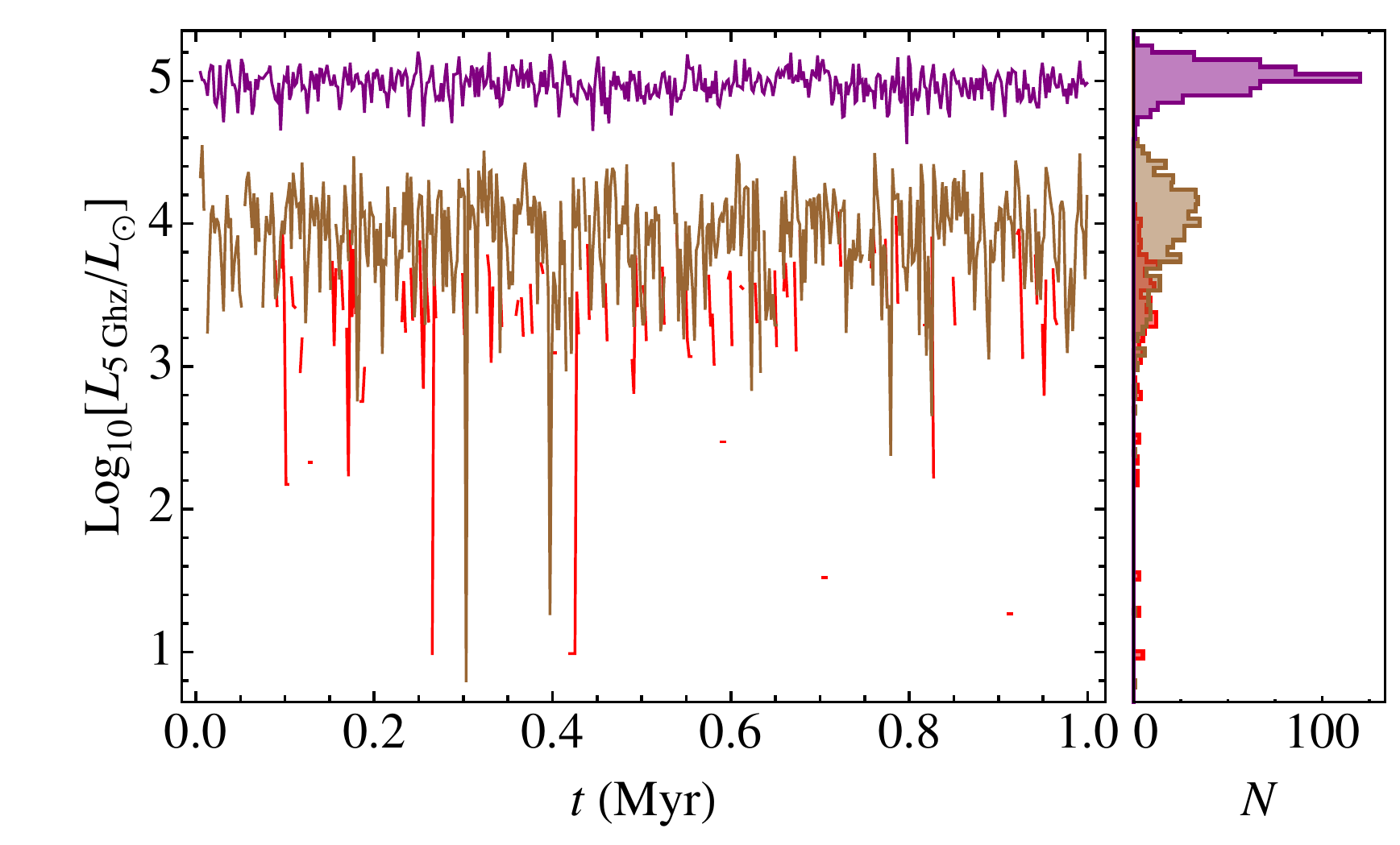}
\caption{Radio luminosity $L$ (Equation~(\ref{eq:lnu})) evaluated for $\nu = 5$~GHz resulting from synchrotron cooling of MS UDRs as a function of tidal disruption rate. In each panel, the three colors correspond to three different tidal disruption rates, with red corresponding to $\Gamma = 10^{-4}$~yr$^{-1}$, brown corresponding to $\Gamma = 10^{-3}$~yr$^{-1}$, and purple corresponding to $\Gamma = 10^{-2}$~yr$^{-1}$. The lines in each panel show $L_{\rm 5 GHz}$ as a function of time $t$ about a single black hole corresponding to our MS case~A (top panel) and MS case~B (bottom panel), with the histograms showing the fraction of time spent at a given luminosity. Giant UDRs (not shown) are qualitatively similar, but generate $\sim 1/3$ the radio luminosity given their smaller kinetic energies on average.}
\label{fig:radio}
\end{figure}

In our own galaxy, the number of active UDRs may only be order unity (e.g. \se), but in other galaxies with potentially larger disruption rates, multiple UDRs may be simultaneously present. Each of these remnants will add to the radio emission emerging from the nuclear cluster, in addition to the SNRs and AGN activity that usually determine a core's radio luminosity. Figure~\ref{fig:radio} shows the radio emission assuming three disruption rates ranging from the fiducial rate $\Gamma = 10^{-4}$~yr$^{-1}$ to $\Gamma = 10^{-2}$~yr$^{-1}$, a rate that may be realized for galaxies with two supermassive black holes in the process of merging \citep{Liu:2013b,Li:2015a,Li:2015b}. Even for non-merging black holes, the stellar disruption rate can approach $10^{-3}$~yr$^{-1}$ for cuspy stellar density profiles \citep{Wang:2004a,Stone:2014a}.

Figure~\ref{fig:radio} demonstrates that even for the fiducial disruption rate there exists a floor radio luminosity of $\sim 10 L_{\odot}$ for case~A, a level comparable to the total radio output of entire non-star-forming galaxies with inactive AGN \citep{Sopp:1991a}. This value is not terribly surprising given that $\tau_{\rm rad}$ (Equation~(\ref{eq:radtime})) is generally longer than the time between disruptions. For case~B, the greatly reduced UDR lifetime results in long periods of no radio output from UDRs (assuming the fiducial rate), punctuated by brief periods of extreme radio luminosities. For both cases, as the disruption rate increases, so too does the radio luminosity, with the luminosity approaching the time-averaged value with increasing numbers of disruptions. For merging black holes with $\Gamma = 10^{-2}$~yr$^{-1}$, the radio luminosity exceeds $10^{3} L_{\odot}$ in case~A, and $10^{5} L_{\odot}$ in case~B, such luminosities rival that of radio galaxies and could be used to infer an enhanced disruption rate.

\begin{figure*}
\centering\includegraphics[width=0.3\linewidth,clip=true]{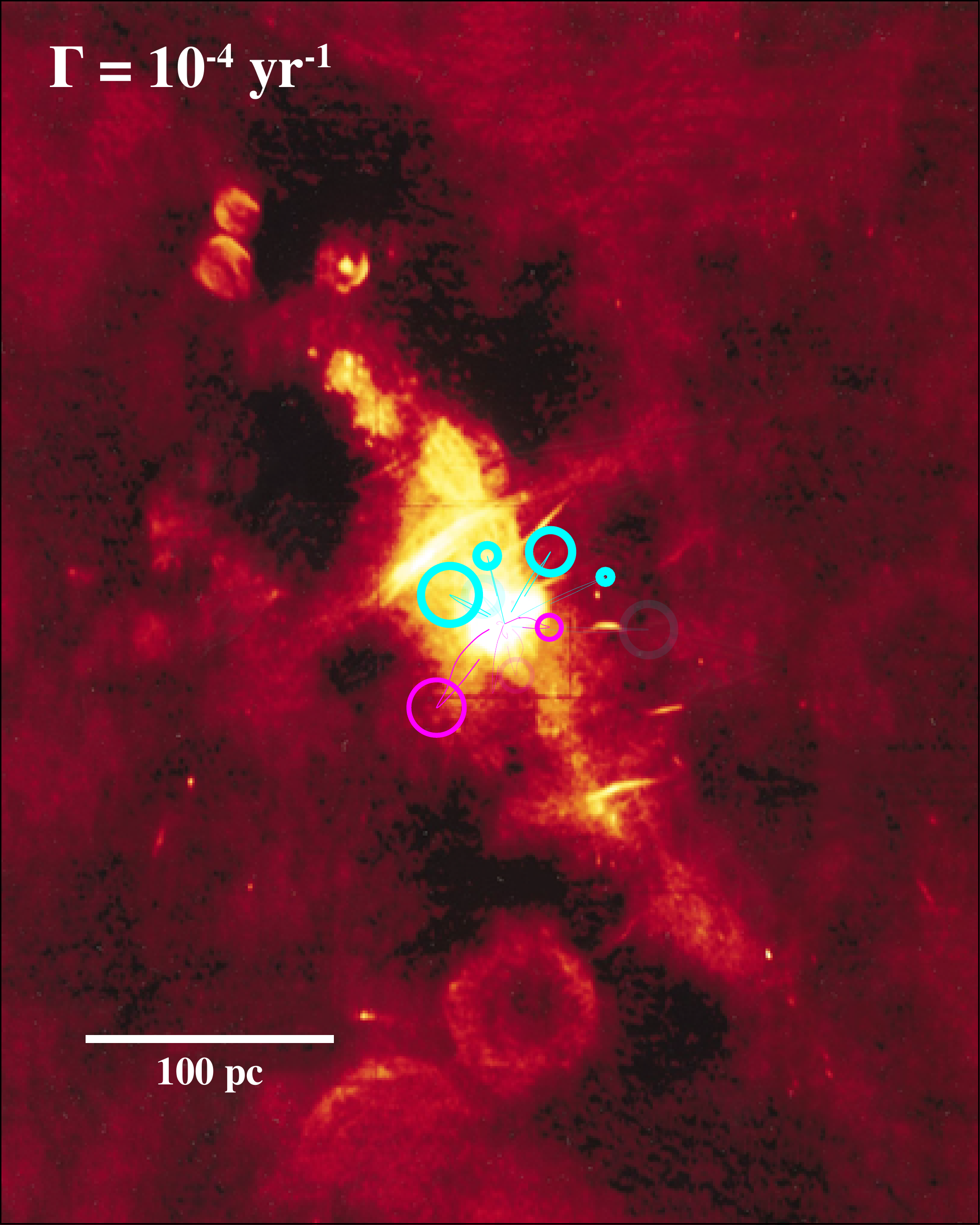}\includegraphics[width=0.3\linewidth,clip=true]{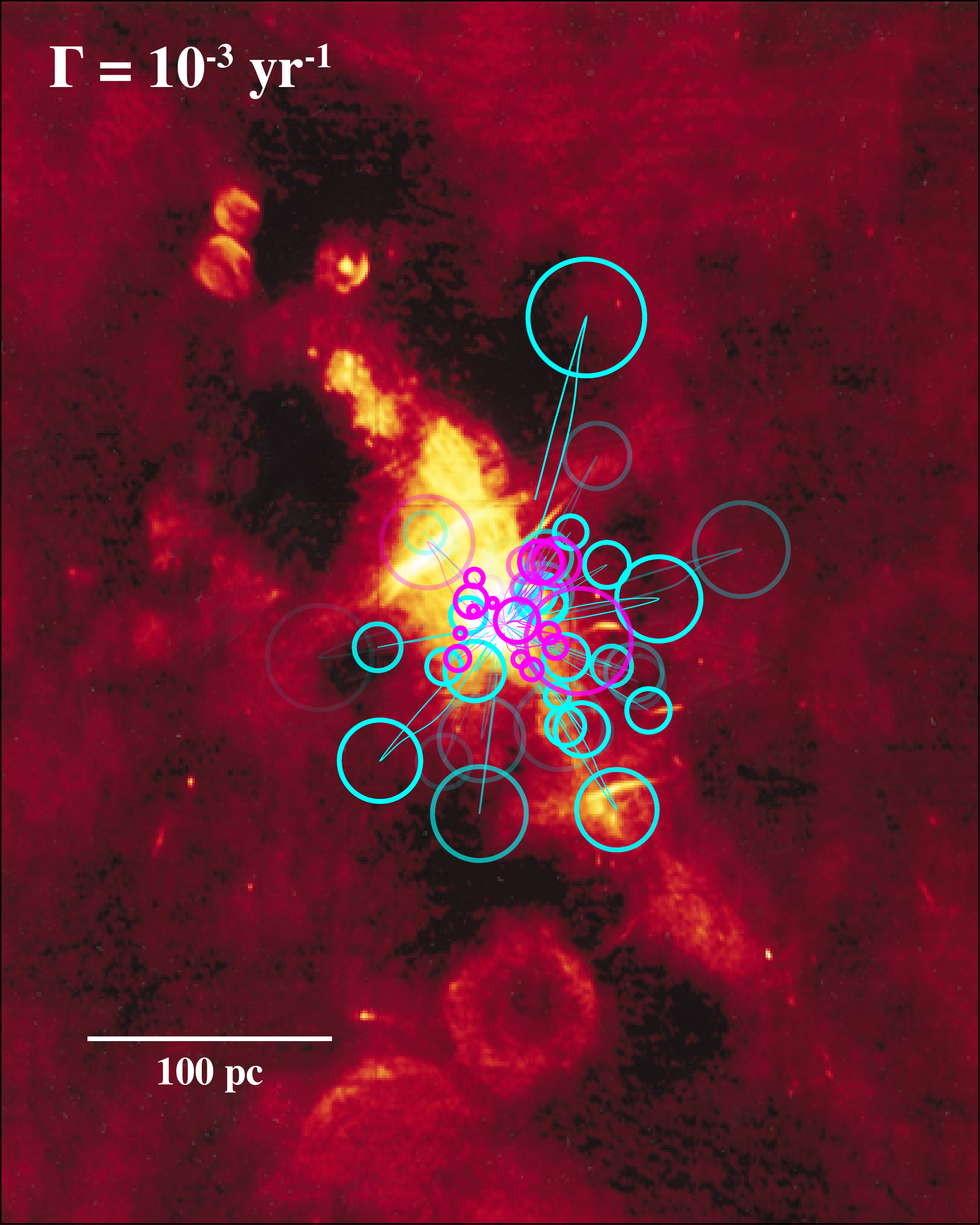}\includegraphics[width=0.3\linewidth,clip=true]{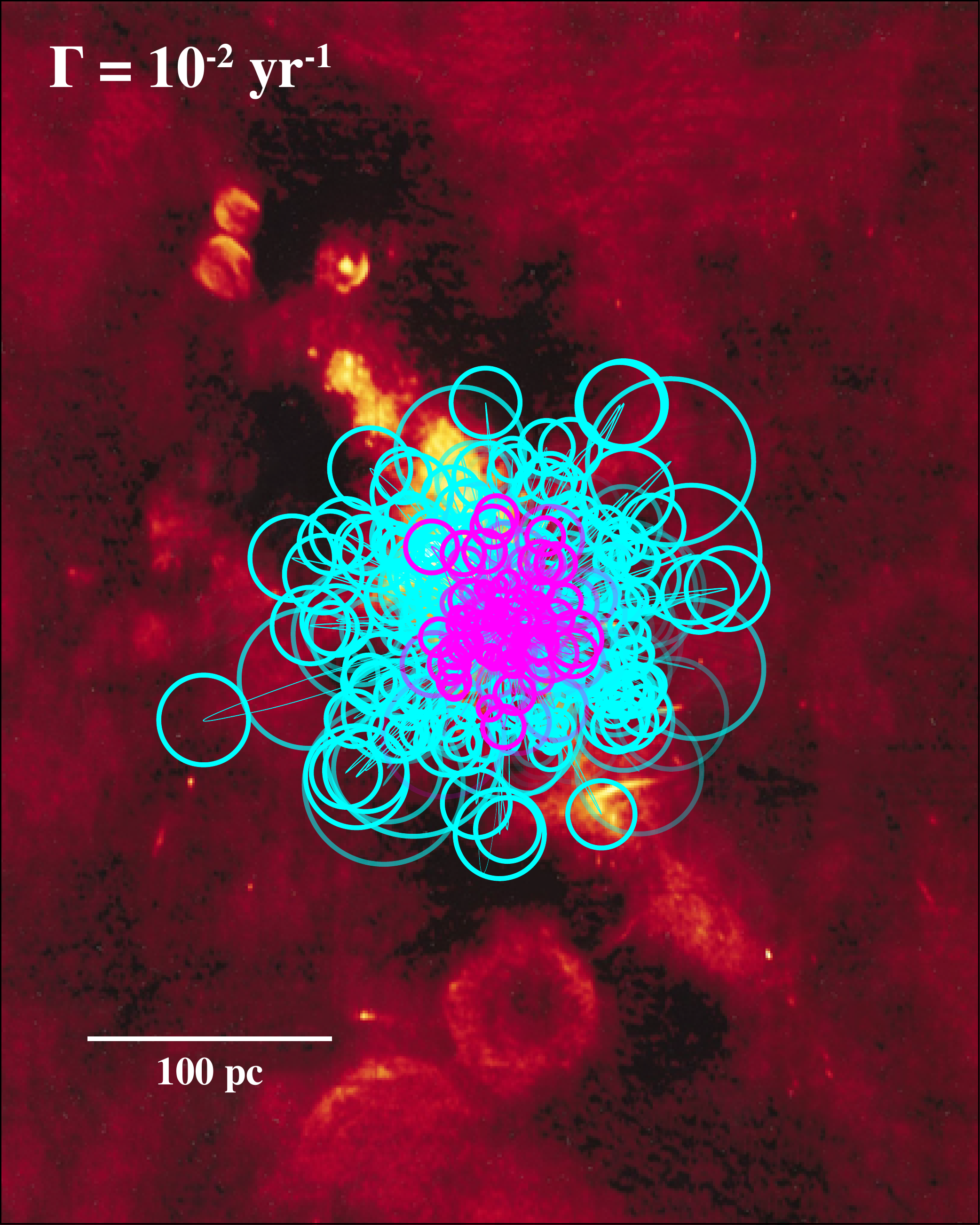}
\caption{Hypothetical distributions of UDSs/UDRs that would be present at the GC for different tidal disruption rates. Each panel shows a randomly-selected snapshot in time of 1,024 UDSs/UDRs using the case~A assumption superimposed upon the wide-field 90 cm VLA image of the GC \citep{LaRosa:2000a}. UDRs/UDSs resulting from MS disruptions are shown as the cyan curves (for the UDSs) and circles (for the UDRs), whereas giant disruptions are shown in magenta. The radius of each UDR is determined by the Sedov solution (Equation~(\ref{eq:rst})) when $t < \tau_{\rm rad}$ and a radiative solution (Equation~(\ref{eq:rrad})) when $t > \tau_{\rm rad}$. When $t = 10 \tau_{\rm rad}$ ($\sim 10^{5}$~yr), we assume the remnant has mixed fully into the ISM and is no longer observable. The disruption rate $\Gamma$ is varied in the three panels, from $\Gamma = 10^{-4}$~yr$^{-1}$ (left panel) to $\Gamma = 10^{-2}$~yr$^{1}$ (right panel), resulting in different numbers of UDSs/UDRs being simultaneously observable. A three-dimensional version of the MS UDS/UDR contribution to the right-most panel of this figure is available at \url{http://goo.gl/7ygeqm}.}
\label{fig:streamsgc}
\end{figure*}

\subsection{Feedback in the central molecular zone}

SNe are one of the dominant drivers of turbulence in star-forming gas, injecting more than enough energy to sustain the turbulent cascade \citep{Nordlund:2003a}. Because they are qualitatively quite similar, the momentum injected by UDSs/UDRs can also act as an important driver of turbulence in the central molecular zone surrounding a supermassive black hole. Figure~\ref{fig:streamsgc} shows the distribution of UDSs and UDRs that could result from tidal disruption rates ranging from the fiducial value to the enhanced rate associated with black hole mergers. Especially at enhanced disruption rates where neighboring UDRs overlap one another, it is clear that UDSs and UDRs can influence the gas dynamics of a significant fraction of the volume surrounding the central black hole, and that the magnitude of the effect is important to calculate.

The UDS will inject its momentum into the surrounding ISM as it stalls in its outward motion; as in the case of a stalled jet, this momentum will be unidirectional, which is unlike a supernova where the net momentum is only a small fraction of its absolute value owing to partial cancellation from the explosion's symmetry \citep[the total net momentum of a SNR is approximately equal to that imparted to its remnant, and is $\sim 10\%$ that of a typical UDS,][]{Wongwathanarat:2013a}. For SNe, the final momentum injected into the ISM by a single SNR is significantly greater than the initial momentum of the remnant as it converts much of its internal energy into kinetic energy by doing $PdV$ work on the surrounding ISM. The rule of thumb is that the momentum injected is given by the initial kinetic energy of the ejecta divided by the velocity of the outgoing shell when the SNR becomes radiative, this is approximately ten times the initial momentum of the ejecta and result in a per-SNR momentum injection of $5 \times 10^{43}$~g~cm~s$^{-1}$ \citep{Kim:2015a}. Utilizing the same logic for UDR, we evaluate Equation~(\ref{eq:vudr}) at $\tau_{\rm rad}$ as determined from Figure~\ref{fig:dc} and evaluate $E_{\rm UDR}/v_{\rm UDR}(\tau_{\rm rad})$; we find that each UDR injects on average $8 \times 10^{42}$~g~cm~s$^{-1}$ of momentum, roughly $1/6$ that injected by a single SNR. As SNe are more frequent than TDEs, UDRs do not contribute significantly to the injection of momentum on galactic scales; however, UDRs are confined to a region within a few tens of pc from the black hole, and thus they may be an important contributor of momentum in this region if the local supernova rate is comparable to the TDE rate. If we assume a disruption rate $\Gamma$, the momentum flux into the ISM from UDRs is
\begin{align}
\dot{p}_{\rm UDR} =~&2 \times 10^{31} \left(\frac{\Gamma}{10^{-4}~{\rm yr}^{-1}}\right) \times\nonumber\\
& \left(\frac{M_{\rm h}}{4 \times 10^{6} M_{\odot}}\right)^{1/3}~{\rm g~cm~s}^{-2}\label{eq:pudr}
\end{align}
There are only a few SNRs located within 100 pc of the GC \citep{LaRosa:2000a}, including \se (which we have argued may be a UDR), suggesting that the local SNe rate is comparable to the tidal disruption rate, as the lifetimes of SNRs and UDRs are similar. Assuming that the SNe rate within this region is $\Gamma_{\rm SNR} = 10^{-4}$~yr$^{-1}$, SNRs inject
\begin{equation}
\dot{p}_{\rm SNR} = 2 \times 10^{32} \left(\frac{\Gamma_{\rm SNR}}{10^{-4}~{\rm yr}^{-1}}\right)~{\rm g~cm~s}^{-2},
\end{equation}
suggesting that SNe likely dominate momentum injection in our own GC despite the rates being similar. However, for galaxies in which the star formation rate is significantly lower than the Milky Way, or for galaxies in which the tidal disruption rate is enhanced, or for more-massive black holes (due to the dependence on $M_{\rm h}$, Equation~(\ref{eq:pudr})), $\dot{p}_{\rm UDR}$ may equal or even exceed $\dot{p}_{\rm SNR}$, and thus provide an important source of feedback for star formation that's not directly tied to the supernova rate.

\section{Discussion}\label{sec:discussion}
In this paper, we have constructed a model for evolution of unbound debris streams (UDSs) resulting from the tidal disruptions of stars by supermassive black holes. We considered both main-sequence and giant star disruptions for two different background density profiles corresponding to the warm ISM and central molecular zones surrounding our Milky Way's central black hole, and found that the resulting UDSs form loop-like shapes that sometimes travel hundreds of pc (in our case~A) before stalling. We calculated the amount of kinetic energy injected by each of these UDS and found this to be significantly smaller than what is quoted in the literature, with a median energy of $E_{\rm UDS} = 10^{50}$~erg for main-sequence disruptions and $10^{49}$~erg for giant disruptions.

We furthermore considered the remnants formed as the result of these UDSs interacting with the background ISM, and concluded that they share many characteristics with the remnants formed from stalled jets and supernovae, and that their dynamics can be well-approximated by the Sedov-Taylor solutions employed to model these other types of remnants. Just like supernova remnants, these unbound debris remnants (UDRs) can accelerate particles that yield cosmic rays and synchrotron emission, which may yield observable emission ranging from radio to gamma rays. By observing radio emission from our own GC, one can place upper limits on the rate of tidal disruptions in the previous few $10^{4}$~yr by counting the number of remnants; Figure~\ref{fig:streamsgc} shows that the number of UDR-like objects in the GC is consistent with a disruption rate that is not greatly in excess of $\Gamma = 10^{-4}$~yr$^{-1}$.

The positive identification of \se as a UDR would have profound implications for the history of the GC. It would imply that there was a tidal disruption event $\sim 10^{3}$ -- $10^{4}$~yr ago, which generated a powerful flare about \sa that would have ionized the surrounding gas, generating a light echo \citep{Ponti:2013a,Ryu:2013a} and even potentially affecting Earth's atmosphere \citep{Chen:2014b}. This accretion may continue to the present day at a low level; if the disruption occurred $10^{4}$~yr ago, an accretion rate with the function form $\dot{M} \propto t^{-5/3}$ would suggest a present-day accretion rate of $\sim 10^{-8} M_{\odot}$ -- $10^{-6} M_{\odot}$~yr$^{-1}$, comparable to estimates for the accretion rate onto \sa \citep{Narayan:1998a,Yuan:2014a}. If G2 also originated from the tidal disruption of a star \citep{Guillochon:2014b}, it is unlikely that G2 and \se originate from the same disruption for the simple reason that the orbit of G2 is incompatible with the location of \se (G2's orbit extends to the southeast of \sa whereas \se lies to the northeast).

Our numerical model, which employed a coupled-differential equation approach to modeling the dynamics of UDR, is simple in its treatment of hydrodynamics, and only considers the first-order drag term associated with a dense body moving obliquely through a low-density medium. We have presumed that the stream's cross-section is initially set by the stream's self gravity, but that recombination eventually causes the stream to widen to the point that it expands ballistically, intercepting a constant solid angle as it travels outwards, all the while remaining cylindrical. Cooling may cause the outgoing stream to fragment into dense clumps separated by lower-density connecting regions \citep{Guillochon:2014b}, and recent simulations \citep{Coughlin:2015a} have suggested that self-gravity may also induce the stream to clump. These clumping effects will at minimum modify the drag coefficient of the outgoing UDSs, potentially enabling them to travel further from the GC before stalling.

Our method for evolving unbound streams could easily be modified to other disruption scenarios such as the tidal disruption of a WD by an intermediate-mass black hole \citep{Rosswog:2009a,Haas:2012a,MacLeod:2015c} or the tidal disruption of a planet by a star \citep{Faber:2005a,Guillochon:2011a}. Additionally, the method could be used to consider non-trivial distributions of matter in GCs that extend beyond the two cases we explored here, as we are free to alter the radial and angular distributions of the ambient gas. As an example, a density distribution that approximates the toroidal structure of the gas in the central molecular zone could be used to better predict the spatial distribution of UDSs/UDRs in our own GC. For both the evolution of the UDSs and UDRs, hydrodynamical simulations in a GC environment are likely necessary to characterize their complete dynamics. It would be especially useful to use such a simulation to model \se to attempt to reproduce the particulars of its morphology, including its ellipsoidal shape and the cannonball, which we have argued could originate from the tip of an outgoing UDS loop.

On the observational side, a robust measurement of Sgr~A~East's composition suggesting a highly-metal-enriched remnant would likely rule out our UDR proposal, as a disrupted star is unlikely to have a metallicity far in excess of the metallicity of the surrounding stars. But as we have argued, there is a high likelihood given the predicted and observed tidal disruption rates that at least one UDR lies in close proximity to our GC, and given \se's observed properties, the object represents a very plausible UDR candidate. Even if \se turned out to be a SNR, the similarity of SNRs and UDRs implies that studying \se would give us valuable clues to how UDRs would evolve in the centers of galaxies, and how they might be detected extragalactically.

\acknowledgements
We thank D.~Finkbeiner, A.~Kamble, A.~Loeb, L.~Lopez, M.~MacLeod, R.~O'Leary, H.~Perets, L.~Sironi, P.~Slane, and A.~Soderberg for useful discussions. This work was supported by Einstein grant PF3-140108 (J.~G.), National Science Foundation grant AST-1312651 and NASA grant NNX15AK81G (M.~M.), CONICYT-Chile through Anillo ACT1101 (X.~C.), and the Gordon and Betty Moore Foundation GBMF-3561 (M.~J.).

\bibliographystyle{apj}
\bibliography{/Users/james/Dropbox/library}

\end{document}